\documentclass[a4wide,10pt]{scrartcl}
\usepackage{graphicx}
\usepackage{amssymb}
\usepackage{amsmath}
\usepackage{multirow}
\usepackage{xcolor,colortbl}
\usepackage[]{natbib}

 \title{Seismic wave propagation in icy ocean worlds}
\author{
Simon C. St\"ahler\thanks{ETH Z\"urich, Institute for Geophysics} \thanks{Leibniz-Institute for Baltic Sea Research, Rostock, Germany}, 
Mark P. Panning\thanks{Jet Propulsion Laboratory, California Institute of Technology, Pasadena, USA}, 
Steve D. Vance\footnotemark[3], \\
Ralph D. Lorenz\thanks{Johns Hopkins University Applied Physics Laboratory, Laurel, USA}, 
Martin van Driel\footnotemark[1], \\
Tarje Nissen-Meyer\thanks{Department of Earth Sciences, University of Oxford, United Kingdom}, 
Sharon Kedar\footnotemark[3]
}

\begin{document}
\maketitle



\begin{abstract}
Seismology was developed on Earth and shaped our model of the Earth's interior over the 20th century. 
With the exception of the Philae lander, all in situ extraterrestrial seismological effort to date was limited to other terrestrial planets. All have in common a rigid crust above a solid
mantle. The coming years may see the installation of seismometers 
on Europa, Titan and Enceladus, so it is necessary to adapt 
seismological concepts to the setting of worlds with global oceans covered in ice.
Here we use waveform analyses to identify and classify wave types, developing a lexicon for icy ocean world seismology intended to be useful to both seismologists and planetary scientists.
We use results from spectral-element simulations of broadband seismic wavefields to 
adapt seismological concepts to icy ocean worlds. We present a concise 
naming scheme for seismic waves and an overview of the features of the 
seismic wavefield on Europa, Titan, Ganymede and Enceladus. In close connection with geophysical interior models, 
we analyze simulated seismic measurements of Europa and Titan that might be used to 
constrain geochemical parameters governing the
habitability of a sub-ice ocean.
\end{abstract}

\begin{figure}
\includegraphics[width=0.9\textwidth]{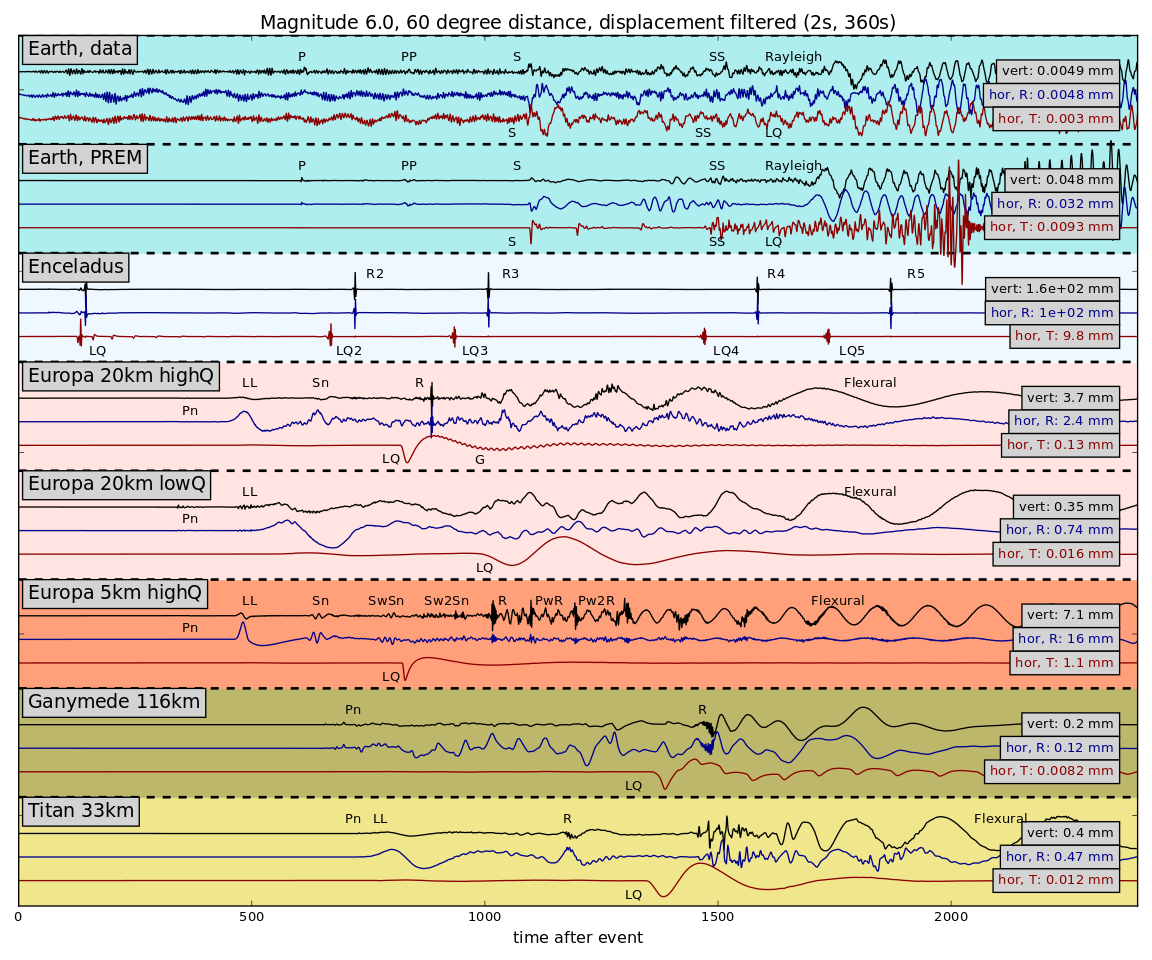}
\caption{Comparison of a measured seismograms on Earth (Earthquake in the Lake Tanganyika region on 2016/02/24, 32 km depth; measured at BFO in Southern Germany in 60 degree epicentral distance) with synthetic seismograms for Earth, Enceladus, Europa, Ganymede and Titan with equivalent angular source-receiver distance. The source model is the real moment tensor, with a source duration of 2 seconds. For the icy moons, the depth has been reduced to 3 km, to ensure that the event is in the ice.}
\label{fig:comparison_moons}
\end{figure}

\section{Introduction}
Most of the larger icy moons of the solar system's gas planets are considered to harbor liquid oceans below their surface \citep{Nimmo2016}. These oceans have been predicted from the estimated internal energy budget by radioactive decay and tides \citep{Lewis1971, Cassen1979} and are very likely to form in a multi-moon situation around a giant planet. In specific settings, even small worlds like Enceladus can have enough energy available to sustain a liquid ocean at least temporarily.
The existence of oceans at Europa, Titan and Enceladus has been identified using density structure inferred from gravitational moment-of-inertia measurements \citep{schubert2004interior}, tectonic interpretations of the ice thickness, and electromagnetic effects consistent with conductive fluids beneath the ice \citep{Kivelson2000,Beghin2010}. Due to the low density contrast between ice and water and ambiguity of electrical conductance measurements, these methods alone cannot constrain the thicknesses of ice and ocean layers uniquely, which are crucial to the ocean chemistry and therefore habitability. Especially in the large worlds Ganymede, Callisto and Titan, high-pressure ice phases at the ocean bottom would could limit water-rock interaction which regulate redox and ocean composition\citep{Vance2014,Vance2017}, with a potentially negative effect on habitability. Briny fluids \citep{Hogenboom1995} or ice convection might mitigate this effect \citep{Choblet2017}, nevertheless the presence, structure and dynamics of these ices is therefore of great scientific interest.

The thickness of a surface ice layer less than a few kilometers can be estimated from radar measurements, as planned for the upcoming ESA JUICE and NASA Europa Clipper missions \citep{Bruzzone2013, Phillips2014}.
Attenuation in the water will make it difficult to measure the ocean depth with radar; seismological measurements would complement the existing methods. Where radar is sensitive to electrical conductivity, and crystal orientation fabric (COF), seismic waves are influenced by elastic parameters and indirectly by material chemistry, COF, and temperature \citep{Diez2013}. A detailed discussion of geophysical measurements linked to habitability can be found in the companion paper by \citet{Vance2017}.

The seismology of an icy ocean world differs from that of Earth. Many of the classical concepts and terminology of seismology become meaningless with the addition of a thick global ocean and ice covering. The global ocean, with its low seismic velocity and absence of S-waves, partially decouples the seismic wavefield in the ice and the underlying solid interior, creating seismic wave types that are not observed at a global scale on Earth. The coming decades will likely bring the first seismological station on an icy moon [e.g.][]\citep{Hand2017}, and mission design begins now. This paper gives an introduction to the general features of icy ocean world seismology, which should help terrestrial seismologists adapt their vocabulary and thinking to this environment.
With the possibility of landers on Europa and Titan in the coming decades, these two worlds are discussed in more detail. Enceladus is also discussed, even though there is currently no strong proposal for a lander and its small, irregular size distinguish it from the other ocean worlds. Ganymede is also discussed as a target for a possible future lander \citep{Vijendran2010}, and as end-member in size and ocean thickness. To keep the paper concise, we will not discuss Callisto and potential ocean worlds in the Uranus and Neptune systems in detail, even though the general concepts presented here apply to them as well. For the four worlds, potential seismic tests for habitability are presented and tested using synthetic seismograms.

Previous studies of seismology on Europa, mostly older than a decade, were therefore limited by the computational methods of the time. The detailed study of \citep{Kovach2001} was based on ray-theoretical methods and analytic derivations and did not calculate any full seismograms. \citet{Lee2003} used ray theory to discuss body waves and surface waves in the ice and ocean layer of Europa, \citet{Panning2006} used the normal mode method \citep{Woodhouse1988} to calculate seismograms, but were limited to long periods (>10s), where some of the ice layer effects are not evident. Computational seismology has progressed rapidly in the last decade \citep[for an overview, see][]{Igel2017}, enabling full modeling of the broadband global wavefield for the first time now. We used the spectral-element solver \textit{AxiSEM} \citep{Axisem}, which is extremely efficient for layered models, since it reduces the 3D wave propagation to a 2D problem. A second companion paper \citep{Panning2017}, couples these results with a seismicity model to estimate the seismic background noise caused by tidal cracking on Europa.

Any realistic lander mission on an icy world will face severe constraints in terms of seismometer mass: Generally, the noise level of a seismometer is controlled by the size of its proof mass: Smaller masses mean higher eigenfrequencies (above 1 Hz) and a higher noise level at long periods. Classical broadband seismometers, like the InSight very Broadband (VBB) main instrument \citep{Lognonne2012,Mimoun2012,Lognonne2016}, have a very good signal-to-noise ratio over frequency range from below 0.01~Hz to 10~Hz, but have total masses of several~kg. Also, their sensitivity to shock and tilted installation impacts mission planning, which is prohibitive as a secondary payload on a lander. Geophones, as proposed for the Mars96 mission \citep{Lognonne1998} are very lightweight, passive instruments, but have a low sensitivity below one Hz and will only record local events and structure. The SESAME instrument on the Philae lander \citep{Seidensticker2007} used piezoelectric sensors, sensitive in the mHz range to see structure on a meter scale around the landing site on 67P/Churyumov--Gerasimenko \citep{Knapmeyer2016}. 
The second instrument on InSight, SP for short period, is a MEMS-based seismometer with a flat response to 120~s period and a noise level slightly higher than that of a Trillium compact borehole instrument on Earth \citep{Pike2016}.
Therefore, the report of the Europa Lander Science Definition Team \citep{Hand2017} defined the baseline instrument as equivalent to the Short Period (SP) instrument of the InSight Mars lander. 


Figure \ref{fig:comparison_moons} shows a seismogram from Earth, compared with synthetic seismograms for an identical source-receiver configuration on Europa, Enceladus, Titan and Ganymede. This comparison highlights the diverse waveforms that are created by the icy-ocean environment. The measured seismogram on Earth is different from the PREM reference synthetic \citep{prem}, since the surface waves traveled along a path through thick continental crust, while PREM contains an oceanic, thin-crust model. No simple global reference model can explain the complete shape of the seismic waveform. Also, ocean-generated microseismic noise appears in the data, with periods of 7 seconds, and another noise signal at around 100s period. This may act as a rough guide for the differences one could expect between simulations and real data from icy worlds.

Seismograms on icy worlds will contain exotic phases, especially the prominent long-period flexural waves in the ice and a strong, non-dispersive longitudinal phase on the R-component (both of which were identified and explained in floating terrestrial ice sheets by \citet{Press1951}). On Enceladus, the Rayleigh (on Z and R) and Love (on T) surface waves orbit the moon in a relatively brief 800 seconds, and they show little energy decrease due to the small size of the object. The seismogram is therefore dominated by its repeated occurrences, which permit determination of the distance between an ice-quake and the receiver \citep[for an application to Mars, see][]{Panning2015}. On Europa, the ice layer is so thin relative to the size of the moon that amplitude and shape of the seismogram depends very strongly on the thickness of the ice layer. On Ganymede, with a 116km ice layer, the overall seismogram shares many features with the Earth: Body waves with low amplitudes arrive first, followed by a complex surface wave train, but very weak exotic phases. 

The paper starts with a description of the models used. In section 3, the seismic wavefield is described, a concise naming scheme is introduced and the various surface wave types in the ice layer are discussed. Section 4 applies the methodology to Europa, Titan, Ganymede and Enceladus, and identifies potential seismic measurements on this representative set of icy ocean worlds.
\section{Models and Modeling}
\subsection{Planetary models}
\sloppypar{
The structural models used here were adopted from the companion paper \citet{Vance2017}. As described in detail there, the models include up-to-date properties for ices, saline oceans, rocky interiors and iron cores, and are thermodynamically self-consistent within the limits of available equations of state.  Radial structures are computed as per \citet{Vance2014}, with self-consistent ice and ocean thermodynamics, using boundary conditions of surface and ice-ocean interface temperature. Thermodynamics for rocks have been added as per \citet{Cammarano2006}. All modeling tools are freely available via GitHub (http://github.com/vancesteven/PlanetProfile).}

Temperature profiles can be tuned to produce different ice shell thicknesses.  It is also possible to consider a range of internal compositions and temperature profiles for the interior below the ocean. We consider the range of plausible ice thicknesses inferred from isostatic surface features, gravitational moment of inertia constraints, and possible radiogenic and tidal heat inputs. 

For basic estimates of attenuation, we followed the approach of \citet{Cammarano2006} to obtain temperature and frequency dependent estimates of shear quality factor, $Q_{\mu}$ with the expression
\begin{eqnarray}
\frac{Q_{\mu}}{\omega^\gamma} &=& B_a\exp\bigg(\frac{\gamma H(P)}{RT}\bigg) \label{anelastic}\\
H(P) &=& g_a T_m,
\end{eqnarray}
in which $B_a=0.56$ is a normalization factor, $\omega$ is the seismic frequency, exponent $\gamma=0.2$ is the frequency dependence of attenuation, and $R$ is the ideal gas constant. $H$, the activation enthalpy, scales with the melting temperature $T_m$ and with the anisotropy coefficient $g_a$, and the values of $g_a$ chosen for various ices are described in \citet{Vance2017}.  The bulk quality factor, $Q_{\kappa}$, is neglected. Values of $Q_{\mu}/(\omega^\gamma)$ in the ice range from $10^7$ in the colder upper regions of the ice, to $10^2$ approaching the freezing point, consistent with measurements in glaciers on Earth \citep{Peters2012}.

\begin{table*}[ht]

\caption{Properties of ocean worlds considered here. 
$^{a}${\citet{schubert2004interior}}
$^{b}${\citet{thomas2010sizes,iess2014gravity}}
$^{c}${\citet{jacobson2006gfs,iess2012tides}} 
$^{d}$\citet{Vance2014}
$^{e}$\citet{Vance2017}
}
\begin{tabular}{|c|c|c|c|c|c|c|c|}
\hline
 & Radius& Bulk density&  Moment of Inertia & ice thickness & ocean depth & high-P ice? \\
 & km &  (kg m$^{-3}$) & & km & km & \\
\hline
Europa$^{a}$	& 1565$\pm8$  & 2989$\pm$46	   & 0.346$\pm0.005 $  	& 5--30	    & >100       & no \\
Ganymede$^{a}$ 	& 2631$\pm1.7$	  & 1942.0$\pm4.8$ & 0.3115$\pm0.0028$ 	& 50--150$^{d}$ & >100$^{d}$ & yes$^{d}$\\
Enceladus$^{b,c}$ & 252.1$\pm$0.2 & 1609$\pm$5     & 0.335 		& 10--55$^{e}$ & 12--50$^{e}$  & no \\
Titan$^{c}$	& 2574.73$\pm0.09$& 1879.8$\pm0.004$ &  0.3438$\pm0.0005$ & 45--120$^{e}$ & >80 & maybe\\
\hline
\end{tabular}
\label{table:planetProps}
\end{table*}

It is important to note that the mineral composition and structure of the rocky interior may be very different from Earth's. For simplicity, we will use the term "mantle" for the rocky layer between water and a potential iron core, which may contain parts compositionally similar to terrestrial crust or even contain unknown minerals created by exotic pressure-temperature regimes. The reader can find relevant literature in the companion paper \citep{Vance2017}. 

\subsection{Wavefield modeling}
This article deals with modeling the full wavefield of a layered, i.e. spherically symmetric icy world. 
To this end, the spectral-element solver AxiSEM was used \citep{Axisem}. It separates the problem of wave propagation in a cylindrically or spherically symmetric object into an analytical solution of the problem in the azimuthal
($\varphi$) direction perpendicular to the source-receiver plane, and a
numerical spectral-element discretisation within the in-plane $r, \theta$, which reduces the numerical cost to that of a 2D method \citep{NissenMeyer2008,Nissen-Meyer2007a} and includes attenuation \citep{VanDriel2014} and anisotropy \citep{VanDriel2014a}. The software was modified to handle general 1D velocity profiles. The changes are included in the recent release 1.4. The properties of the SEM meshes and the resulting numerical cost are shown in appendix \ref{app:sim_properties} Compared to the typical meshes for global wave simulation on terrestrial planets, smaller elements are needed due to the low wave velocities in the ocean. As these are fluid elements, the maximum allowed time step is larger than on earth, so that the computational cost is about a factor 2 larger than for an terrestrial planet of similar radius.

The reciprocity of the Green's function permits switching of the locations of source and receiver of a seismic wavefield. This has been used in the Python package \textit{Instaseis} \citep{vanDriel2015}, which allows to reconstruct seismograms for arbitrary source and receiver locations from a precalculated wavefield database. Instaseis uses the stored displacement wavefield to calculate strain, which allows to simulate arbitrary moment tensor and single force point sources with arbitrary source time functions. 
Theoretical arrival times were calculated with TauP software package \citep{Crotwell1999} in the implementation of ObsPy 1.0 \citep{Krischer2015}.

To estimate the effects of three-dimensional heterogeneities and scattering, we introduced lateral heterogeneities into some runs. Since AxiSEM assumes an axially-symmetric model, these heterogeneities have to be symmetric around the source axis. This creates a limited scattering effect, especially since no off-path scattering is simulated, but it serves as a good first-order approximation to estimate the total effect of heterogeneities. For the scattering, we implement a von K\'arm\'an random medium with a correlation length of 5~km and velocity variations of 10\%. This means that the correlation length is similar to the shortest wave length of P-waves. In models with lateral heterogeneities to simulate scattering, these structures become symmetric around the receiver axis.

Input models for all planets discussed in this article are available as an electronic supplement. The seismic wavefields for the models discussed here are available as Instaseis databases on http://instaseis.ethz.ch/icy\_ocean\_worlds/ and can be used freely for further analysis. Waveform databases for Earth models can be accessed from the IRIS \textit{Syngine} service \citep{Krischer2017}.
\begin{figure}
  \centering
  \begin{tabular}{ccc}
    \hspace{0.5cm} Europa \hspace{1.8cm} & \hspace{0.5cm} Ganymede & \hspace{1.8cm}\hspace{0.5cm} Enceladus 
  \end{tabular}\\
  \includegraphics[width=0.9\textwidth]{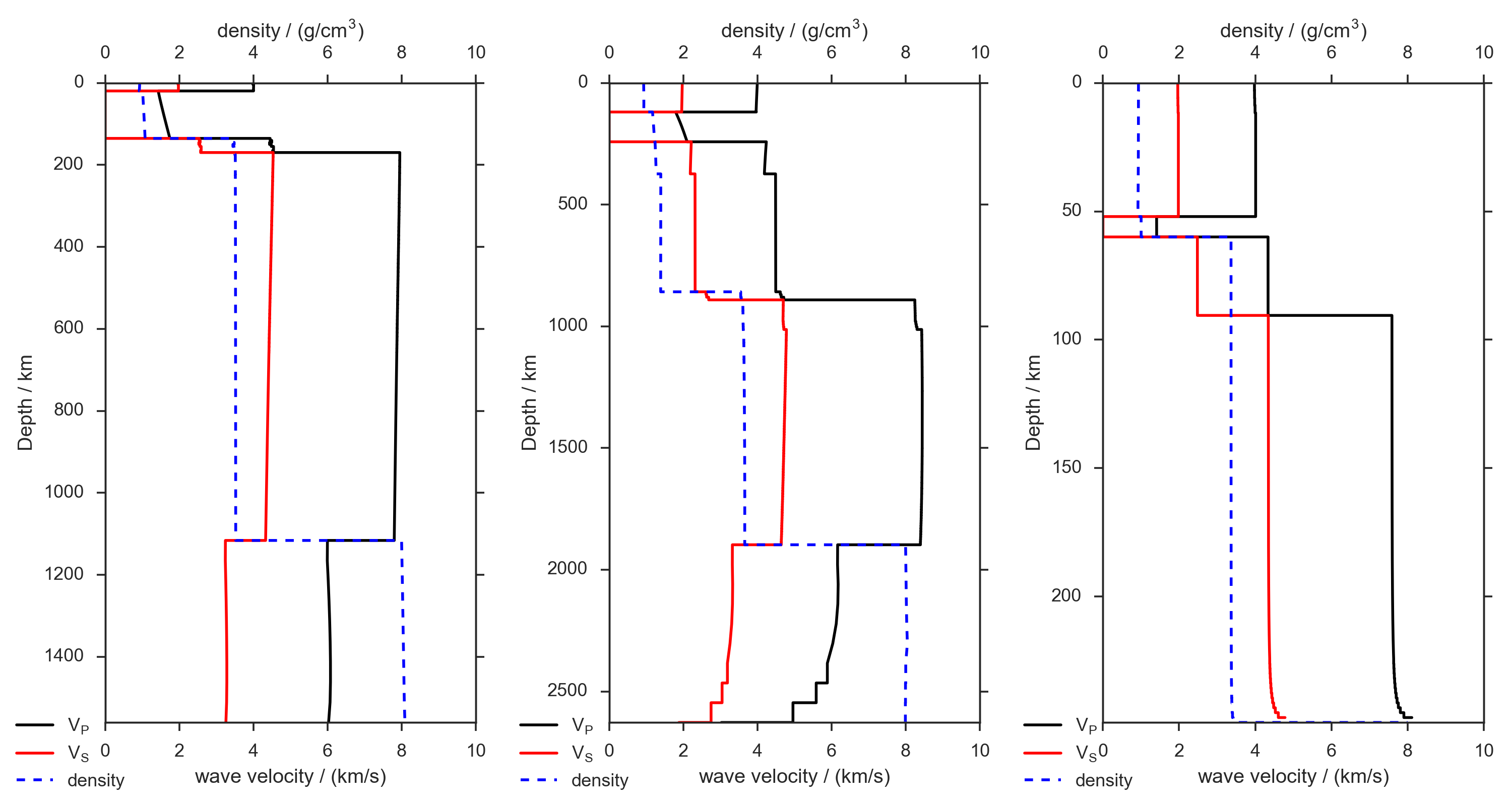}
  \caption{Velocity Profiles of three icy moons \citep[electronic supplement]{Vance2017}: Europa (left), with a 20~km thick ice layer on top of a 116 km deep 
           ocean. Ganymede with (center) with a 124 km thick ice layer on top of a 116 
           km deep ocean and 600 km of high-pressure ice phases. Enceladus (right) has a 52 km ice layer on top of 
           an extremely shallow ocean and no core.
           Compared to Earth, the velocity gradients within each layer are very small due to the relatively low 
           pressure gradients. The only considerable velocity gradients exist in the ocean.}
  \label{fig:velocity_models}
\end{figure}

\section{The seismic wavefield}
\subsection{Phase naming}
\begin{center}
\begin{tabular}{|p{3.1cm}p{8.9cm}|} 
\multicolumn{2}{p{8cm}}{\textbf{Body waves}}\\ \hline
P & P-wave in upper ice layer or mantle\\
S & S-wave in upper ice layer or mantle\\
F & P-wave in \textbf{F}luid layer\\
K & P-wave in the (solid) core
\\ \hline
\multicolumn{2}{p{12cm}}{\textbf{Boundaries}}\\ \hline
 & \textit{Surface reflections are not specifically named}\\
e & Bottom of the ice (\textbf{e}is) layer\\
o & \textbf{O}cean floor\\
g & Top of the rocky layer (\textbf{g}ranitic crust), if high pressure ice is present below ocean \\
m & Top of the \textbf{m}antle, if extra layers are present between ocean and mantle\\ \hline
\multicolumn{2}{|p{12cm}|}{Interface interactions}\\ \hline
$^x$/\textasciicircum\textit{x} & Underside reflection on interface \textit{x}\\
\textit{x} & Topside reflection on interface \textit{x}\\ \hline

\multicolumn{2}{p{12cm}}{\textbf{Specific phases}}\\ \hline
Pe/Se & P or S-waves bottoming in the uppermost ice layer\\ 
PeP/SeS & P or S-waves being reflected at the ocean-ice boundary\\
Pn/Sn & "\textbf{n}ormal" P- or S-waves traversing all water layers, turning upwards in the mantle\\
PoP/SoS & P or S-waves being reflected at the ocean floor\\
PmP/SmS & P- or S-waves being reflected at the top of the \textbf{m}antle\\
\hline
\multicolumn{2}{p{12cm}}{\textbf{Prefixes for multiples.} Append by integer \textit{N} for multiple reverberation. \textit{y} is a wildcard for any following phase.}\\ \hline
Pf\textit{Ny} / Sf\textit{Ny} & P or S-waves being reflected \textit{N} times at the ocean floor and at the ocean/ice boundary; reverberation in the \textbf{f}luid ocean\\
Pe\textit{Ny} / Se\textit{Ny} & P or S-waves being reflected \textit{N} times at the ocean/ice boundary and at the surface; reverberation in the ice (\textbf{e}is)\\
\hline
\multicolumn{2}{p{12cm}}{\textbf{Surface waves}. \textit{N} is integer and indicates wave packets traveling along the minor arcs (odd numbers) or major arc (even numbers) of the great circle}\\ \hline
C\textit{N} & \textbf{C}rary wave in the ice layer\\
R & \textbf{R}ayleigh wave at the surface\\
Ro & \textbf{R}ayleigh-like wave on the \textbf{o}cean-floor: Scholte wave\\
Re & \textbf{R}ayleigh-like wave at the ice (\textbf{e}is)-ocean interface\\
LF & \textbf{L}ong-period \textbf{F}lexural wave, in the ice layer\\ 
LL & \textbf{L}ong-period \textbf{L}ongitudinal wave, in the ice layer\\
LQ & \textbf{L}ong-period Toroidal mode (\textbf{Q}uerwelle), in the ice layer\\
G & Love wave, in the ice layer\\
Go & Love wave, in the ocean floor\\
\hline

\multicolumn{2}{c}{}
\label{tab:phase_naming}
\end{tabular}
\end{center}
To facilitate seismic analysis, it is necessary to be able to name phases (i.e. distinct signals in the seismogram) unambiguously. The terrestrial naming scheme adopted by IASPEI \citep{Storchak2003} needs to be extended in the context of icy moons to account for the various ice phases. A completely non-ambiguous scheme, which marks every interface crossing, was proposed by \citet{Knapmeyer2003}, but since it produces rather cumbersome phase names, we propose a simplified version here. \citet{Lee2003} proposed a scheme specifically for local waves, but since it cannot be generalized to global wave propagation, we decided not to use it here. 
In our scheme, all names from the IASPEI scheme keep their meaning, where applicable, but we add a few new letters for ice-specific phases. Because ``i'' and ``I'' have already been used for inner core phases, we propose ``e'' as the letter generally related to ice phases, using the German word \textit{Eis} (analogous to core phases being called ``K'' from \textit{Kern}). 

\begin{figure}
\includegraphics[width=0.5\textwidth]{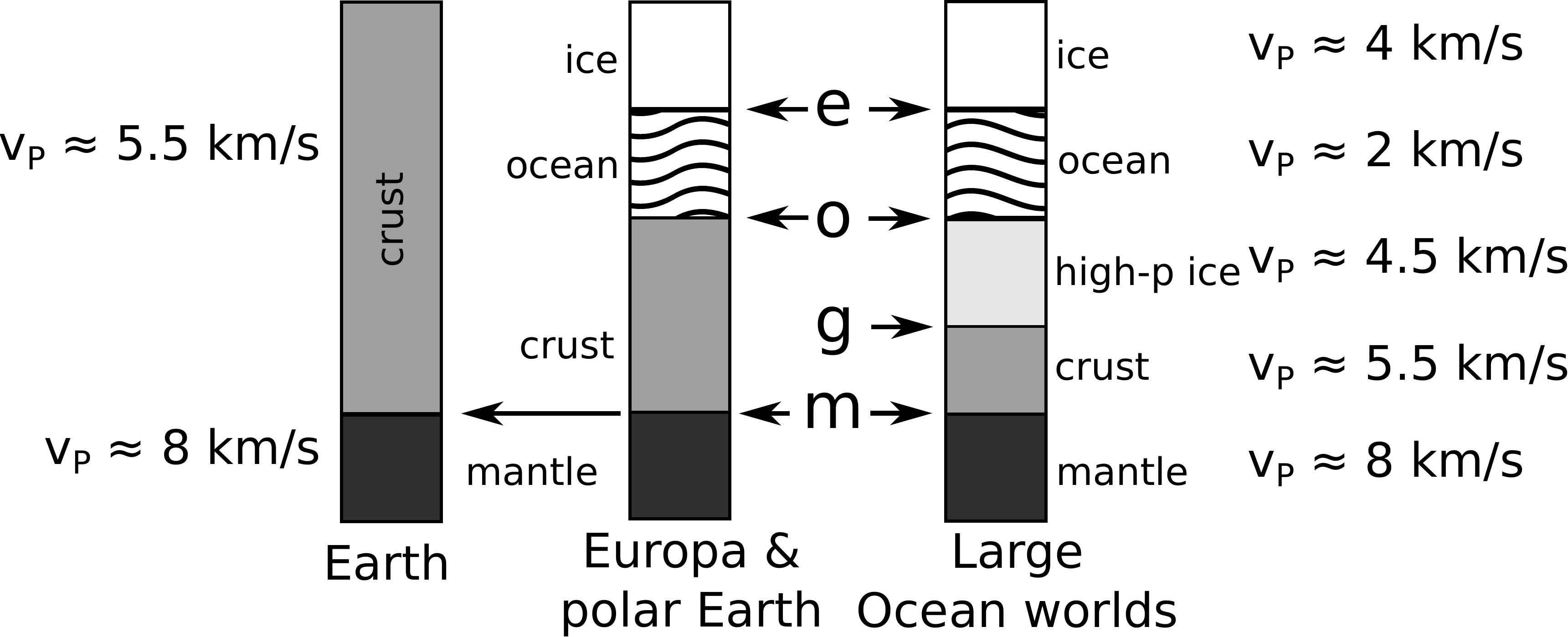}
\caption{Comparison of the structure of the uppermost layers in Earth and an ocean world. The letters ``e'', ``o'', ``g'', ``m'' mark the interfaces in seismic phase names. Note that sea ice in the polar regions of Earth qualifies as an ocean world in this definition.}
\label{fig:crustal_struct}
\end{figure}

The main issue at hand is that there is only the Moho as a strong, global near-surface interface on Earth, but there are at least 3 in ocean worlds (a Moho, plus the ice-bottom and the ocean floor and possibly an interface between high-pressure ice and the crust). Standard IASPEI nomenclature would denote these interfaces by their depths. Since this depth is poorly constrained on ocean worlds, we propose to give these interfaces descriptive letters: ``e'' for the bottom of a surface ice layer, ``o'' for the bottom of an ocean, ``g'' for the top of the rocky layers and ``m'' for the top of the mantle. To use this scheme for glaciers on Earth or worlds like Ceres, the bottom of the ice layer should also be called ``e'' if it is in contact with a solid rocky interior, see \ref{fig:crustal_struct} for an overview.

The IASPEI standard for marking reflections on these interfaces (appended ``+'' or ``-'') is not widely used in the community, so we stick to the nomenclature of \citet{akirichards}, where topside reflections are noted by the interface descriptor (or depth), as in PeP, and underside reflections are noted by the descriptor as a superscript (e.g. P$^{\mathrm{m}}$P). Where the usage of a typographic feature is problematic, the mnemonic TauP convention of P\textasciicircum{}mP should be used instead.
\begin{figure}
\includegraphics[width=0.5\textwidth]{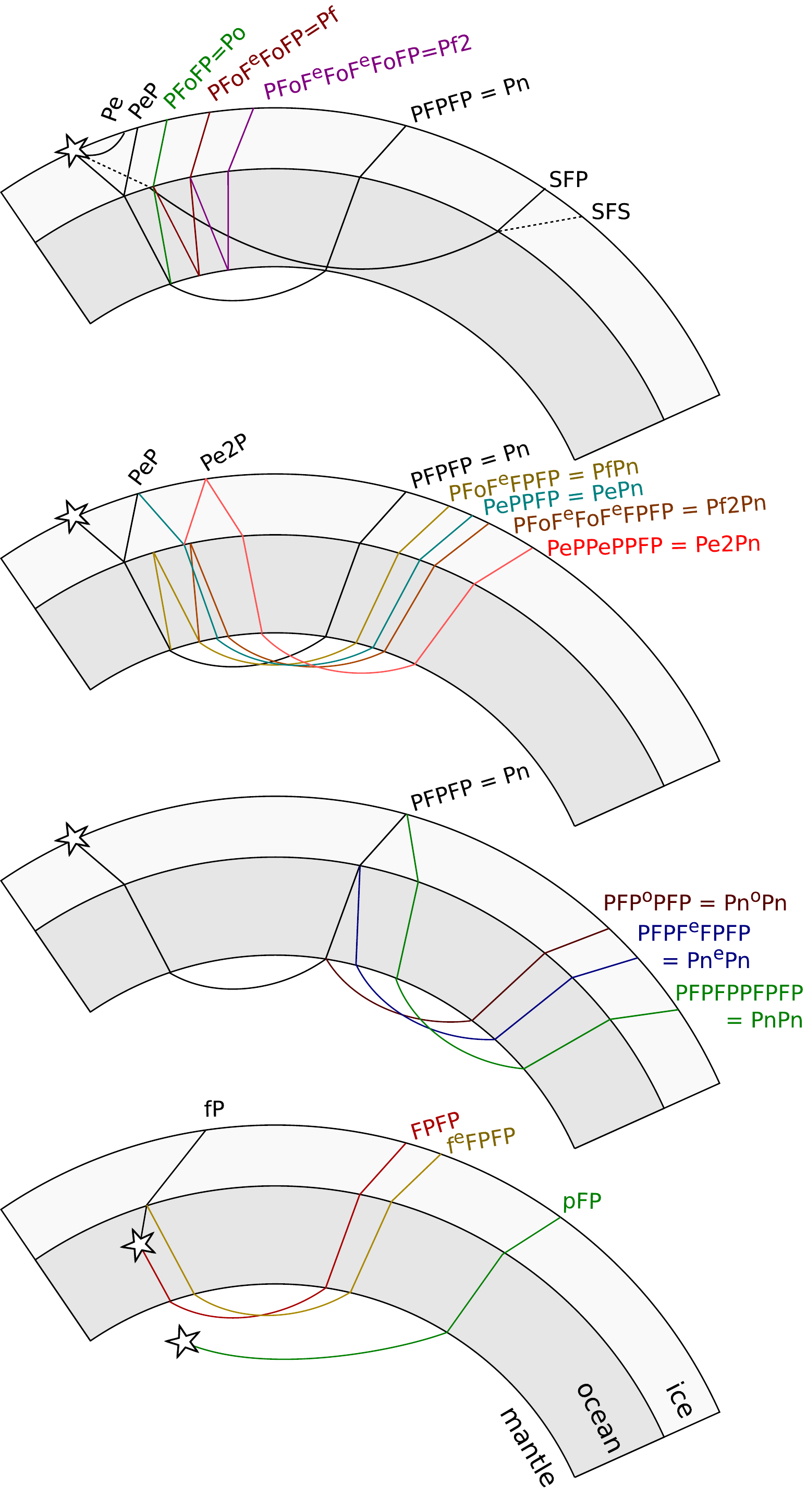}
\caption{Paths of various P-phases, including multiples from the ice layer, the water layer and surface reflections for a Europa-like icy moon. Note that the thickness of the ice layer is not to scale.}
\label{fig:paths}
\end{figure}

Terrestrial seismology calls compressional waves in the ocean either "H", for a hydroacoustic wave from a source in the water, or "T", for a hydroacoustic wave from a source in the ground. Both wave types are linked to the SOFAR channel of low sound speed between 500 and 1500 meter depth in Earth's ocean \citep[SOund Fixing And Ranging;][]{Tolstoy1950, Ewing1951}. This channel is an effect of the specific contributions of depth-dependent salinity and pressure to sound speed in earth's ocean and most likely does not exist in ocean worlds. We therefore propose to mark a leg of the path that traverses the ocean generally with ``F'' (\textbf{f}luid). The terrestrial ``H''-wave would then be ``FP'' and the ``T''-wave ``pFP''.

Waves in the uppermost ice layer and in the mantle are both called P or S, depending on their type. 
A PFPFP wave is therefore one that started within the ice layer, crossed 
the ocean, traveled through crust and potentially mantle, turned there and went back through the 
ocean to the surface. To retain a certain elegance in naming phases, we propose to use ``Pn'' for this wave type, analogous to the mantle phases on Earth as proposed by \citep{Mohorovicic1909}. To specifically highlight local phases that never left the ice, we propose to add an ``e'': ``Pe'', ``Se''. Note the distinction from ``PeP'', which was reflected at the ice bottom.

To keep phase names concise, we introduce a few more abbreviations: ``Po'' is short for a topside reflection from the ocean floor (full: PFoFP), ``Pm'' is short for a topside reflection from the mantle top (full: PFPmPFP). ``Pf'' is a single reverberation within the water (PFoF\textsuperscript{e}FoFP). These abbreviations can also be used as prefixes to mark multiples on other phases, as in ``PfPn'' = ``PFoF\textsuperscript{e}FoFPFP''.
See fig. \ref{fig:paths} for an overview of ice- and ocean-interacting body wave paths. As on Earth, a number following one descriptor means a multiple reflection. ``Pf2Pn'' was reflected twice at the ocean bottom and twice at the bottom of the ice layer before traveling as a P-wave through the mantle.

Up-going seismic waves from an event in or below the water layer should be distinguished. This differs from the terrestrial usage, where source depth is explicitly not coded in the phase name. We therefore propose to write the first letter in lowercase for events below the ice, e.g. pFP for an event originating below the ocean layer or fP for an acoustic source in the ocean. This is a slight extension of the terrestrial usage, where only rays with an upward leg are noted as such. This naming scheme combines the brevity of the IASPEI scheme with the flexibility of Knapmeyer's scheme, while at the same time not being limited to local waves, as in the scheme by \citet{Lee2003}. See table \ref{tab:name_examples} for examples of phase names in our scheme, Knapmeyer's scheme and the input to the popular TauP software.

\begin{figure}
\includegraphics[width=0.9\textwidth]{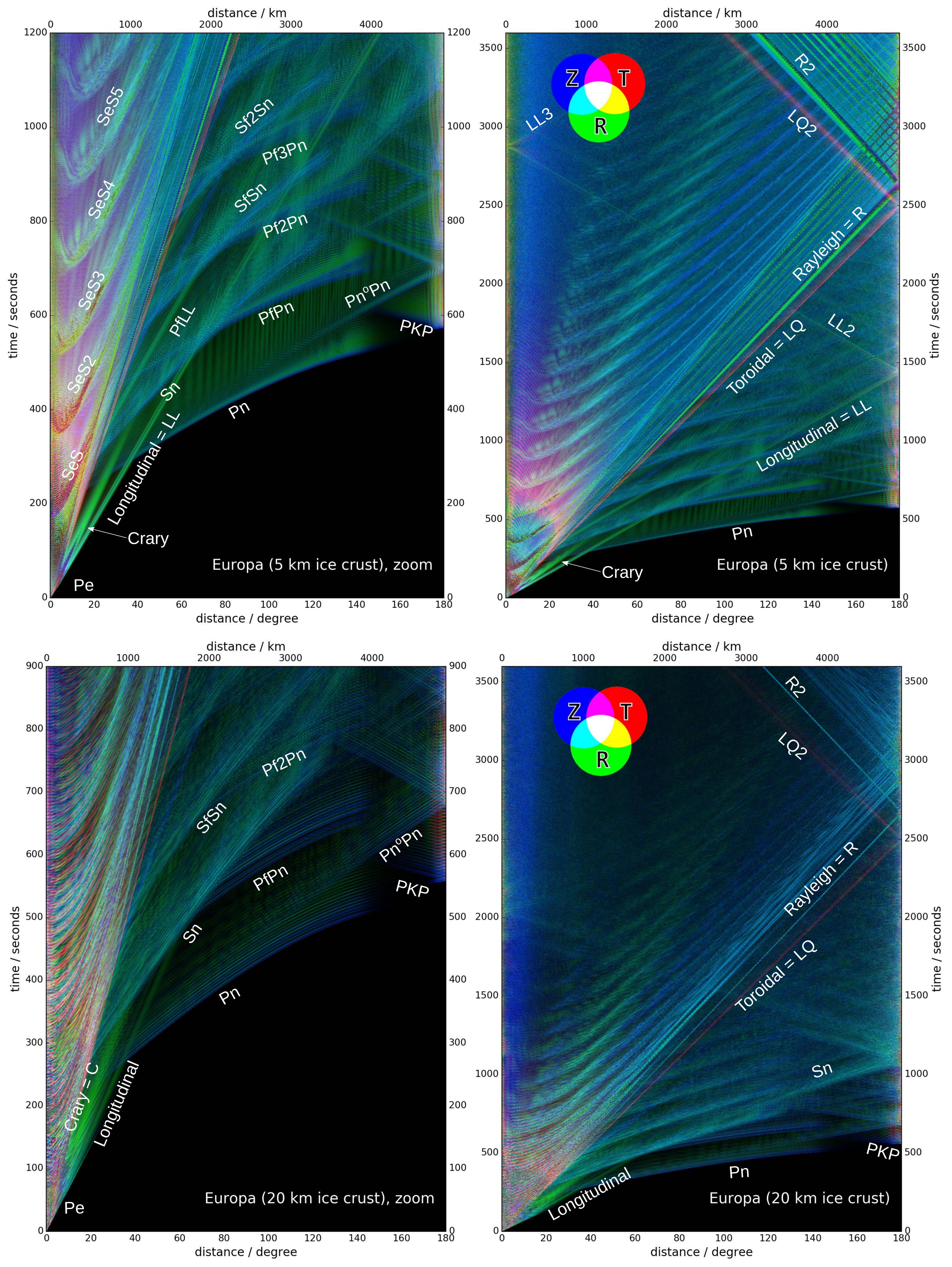}
\caption{Global stack of seismograms \citep{Astiz1996} with phase names according to the scheme in table \ref{tab:phase_naming} for a Europa model with 5~km ice thickness (top) and 20~km ice thickness (bottom). Vertical motion is blue, transverse red and radial green.}
\label{fig:global_stack}
\end{figure}

\begin{table}[ht]
\begin{tabular}{p{2cm}cccc} \hline
Our scheme (short) & our scheme (full) & Knapmeyer & TauP & Lee\\ \hline
Pn & PFPFP & POPoPopOp & P20P136P136P20P & N/A\\
Pe & P & P & P & P\\
PoP & PFoFP & POPo+POP & P20Pv136P & PCP\\
PnPn & PFPFPPFPFP & POPoPopOpPOPoPopOp & P136PP136P & N/A\\
PFSFP & PFSFP & POPoSopOp & P20P136S136P20P & N/A \\
Pf & PFoF$^{\mathrm{e}}$FoFP & POPo+pO-Po+pOp & P20Pv136p\textasciicircum20Pv136p20p & N/A \\
C & C & N/A & 2950kms & N/A 
\end{tabular}
\caption{Examples of phase names in our proposed naming scheme, the scheme proposed by \citet{Knapmeyer2003} and the one proposed by \citet{Lee2003}. The latter is only applicable to local phases that did not enter the rocky part below the ocean. Surface-wave-like phases like the Crary phase cannot be named in any of the other schemes, only TauP can describe it by its phase velocity.}
\label{tab:name_examples}
\end{table}
We show a visualization of the global wavefield in fig. \ref{fig:global_stack} that follows the example of the IRIS global stack \citep{Astiz1996}. Seismograms for different distances are color coded, where vertical motion is color coded blue, motion in R direction (away from the event) in green and transverse motion in T. Each trace is normalized separately. We see that in the coda of Pn-waves (see also fig. \ref{fig:Pcoda}), every body wave phase is followed by multiples from the mantle-ocean and ocean-ice boundaries. 
\subsection{General features of the icy-moon wavefield}
\label{sec:general_features}
\begin{figure}
  \centering
  \includegraphics[width=0.9\textwidth]{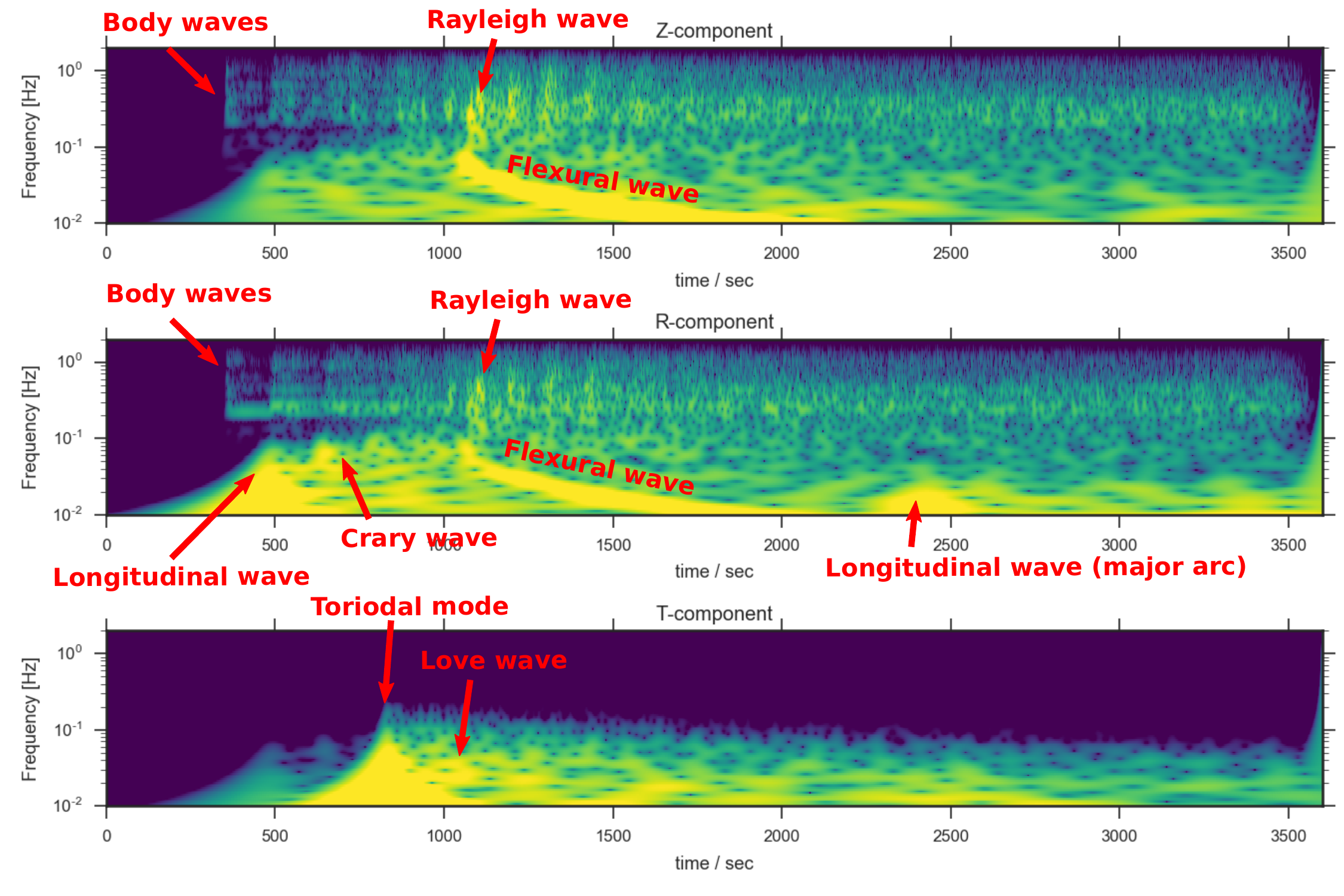}
  \caption{Spectrogram of an event in a Europa model with 20~km ice thickness in 60 degree distance, showing the different spectral characteristics of the wave types. Even though the flexural wave and the longitudinal wave dominate a broadband seismogram (see fig. \ref{fig:comparison_moons}), other phases can be identified, because they are occur at higher frequencies.}
  \label{fig:spectrogram}
\end{figure}
\subsubsection{Seismic phases in the ice}
Seismic events within the ice layer will create a range of seismic wave types that are not observed in global seismograms on Earth (see fig. \ref{fig:spectrogram} for a spectrogram, which separates the spectral characteristic better; see Table \ref{tab:wave_types} for a summary of the different wave types, including their group and phase velocities.).  This is especially the case for ice layers thinner than 40~km, as expected for Europa. The three most prominent new phases are flexural waves, longitudinal waves, and the Crary phase. As shown by \citet{Press1951}, the response of a floating ice layer of thickness $d$ is similar to that of an infinite slab. \citet{Kovach2001} applied this to the situation on Europa, so that we can summarize the most important results here, assuming a compressional wave velocity $v_{\mathrm{P, ice}}=4~\mathrm{km/s}$ and a shear velocity $v_{\mathrm{S, ice}}=2~\mathrm{km/s}$:
\begin{description}
\item[Rayleigh waves:] Short period surface waves with vertical/radial polarization do not see the bottom of the ice layer and can therefore be described as a Rayleigh wave with the usual properties, if the inverse wavenumber $k^{-1}$ is smaller than the ice thickness $d$: ($kd = 2\pi\frac{d}{\lambda} > 1$). \\
The velocity gradient with depth is very small in the ice, so Rayleigh waves have little dispersion. For small wavelengths, the usual result of a phase velocity $c_R=0.92 v_{\mathrm{S, ice}}\approx 1.8~\mathrm{km/s}$ is valid. 
\item[Rayleigh wave (ice bottom):] At short periods, an interface wave travels at the ice-water interface, called \textit{Stoneley} wave by \citet{Press1951} and \citet{Kovach2001} (even though that term usually describes solid-solid interface waves). An interesting property is that its maximum phase velocity is controlled by the sound velocity \textit{of the ocean}: $c_{\mathrm{St}} = 0.87v_{\mathrm{P, water}}\approx 1.3 \mathrm{km/s}$. Its signal is only detectable for ice shells of a few kilometers, but there, the arrival time difference from the surface Rayleigh wave constrains the sound velocity of the ocean.
\item[Flexural waves:] Long-period vertically polarized surface waves, with $kd<1$ are perturbed by the bottom of the ice layer and travel as full-layer flexural waves. The most interesting result is an intrinsic dispersion $c_{\mathrm{Fl}} \propto \sqrt{f}$ that is independent of the velocity gradient in the ice. Due to this frequency dependence, the short periods arrive first, and the group velocity of long-periods goes to zero. The cutoff frequency is where the Rayleigh regime (see above) starts. \citet{Panning2006} showed the effect of different ice thicknesses on this wave train.
\item[Longitudinal wave:] Another long-period solution is a wave with motion in the radial direction, in phase over the whole ice layer, described analytically in \citet{Press1951}. This wave has a phase and group velocity defined by the expression
\begin{equation}
v_L=2v_S\sqrt{1 - \left(\frac{v_{\mathrm{S}}}{v_{\mathrm{P}}}\right)^2},
\end{equation}
with $v_L\approx3.46$ km/s for the example P and S velocities. It therefore travels as a fast, non-dispersive pulse along the surface. This means that it arrives much before all other ice phases at long distances, and even before mantle body waves at intermediate distances (see tab. \ref{tab:distance_regimes}). It is the most prominent feature in the broadband seismogram. Both flexural waves and longitudinal waves are solutions for wavelengths long relative to the ice shell thickness, and so only show up at relatively low frequencies ($kd<1$).

\item[Love waves:] 
  No shear waves exist in the fluid ocean, so the ice layer is a wave guide for SH waves, which are completely reflected at surface and ice-ocean interfaces. This results in a dispersive group velocity of 
  \begin{equation}
  U(f) = \dfrac{v_{\mathrm{S, ice}} }{\sqrt{1 - \left(\dfrac{n v_{\mathrm{S, ice}} }{2fd}\right)^2}},
  \end{equation}
  where $n=1,2,3,...$ is the mode of the Love wave, $v_S$, the shear wave speed and $d$ the ice thickness. Similar to the flexural wave, we have an inverse dispersion, where infinite frequencies travel with $U=v_{\mathrm{S, ice}}$, while lower frequencies are slower.
  The minimum frequency of the $n$-th mode Love wave is 
  \begin{equation}
  f\mathrm{min} = \frac{n v_{\mathrm{S, ice}}}{2d} \approx \frac{n}{d} \mathrm{mHz},
  \end{equation}
  for ice with $v_{\mathrm{S, ice}}=2~\mathrm{km/s}$. This means that $f_\mathrm{min}=0.2$~Hz for the fundamental mode in a 5~km ice shell and 0.05 Hz for a 20~km ice shell. It thereby constrains the minimum ice thickness along the path. 

\item[Toroidal wave:] For long periods below the minimum frequency of the guided Love wave, the whole ice layer moves in phase. Analogous to the longitudinal wave, this results in a non-dispersive, low-frequency wave with high amplitude and a phase and group velocity equal to the SH velocity of the ice shell.
  
\item[Crary phase:] As a result of SV waves being completely reflected at critical angle for SV-P conversion on the boundaries of the ice, another phase arises with interesting properties \citep[p. 282]{Kovach2001}: As first described by \citet{Crary1954}, with application to terrestrial floating ice in the Arctic, it has mainly radial displacement, a phase velocity of $v_{\mathrm{P}}$ (for homogeneous ice layers) and a harmonic frequency spectrum characterized by 
\begin{equation}
f_{\mathrm{Cr}} = \dfrac{(n+1)~v_{\mathrm{S}}}{2d \sqrt{1 - \left(\frac{v_{\mathrm{S}}}{v_{\mathrm{P}}}\right)^2}},
\label{eq:Crary_frequency}
\end{equation}

where $n=1$ defines the characteristic frequency, and larger integer values of $n$ representing overtones.
Its group velocity is $U\approx0.75v_{\mathrm{P}}$, which means it arrives slightly after the longitudinal wave. Measurement of $f_{\mathrm{Cr}}$ constrains the ice thickness $d$, if $v_{\mathrm{S}}$ and $v_{\mathrm{P}}$ are known. Since $f_{\mathrm{Cr}}(d=5\mathrm{km}) = 0.46~\mathrm{Hz}$ and $f_{\mathrm{Cr}}(d=20\mathrm{km}) = 0.11~\mathrm{Hz}$, the thickness of Europa's ice layer can be measured with a seismometer that is sensitive at these frequencies. 
For ice thicknesses above 40~km, the maximum frequency becomes too low to be observable for realistic events with a space-ready seismometer.
\end{description}

Especially for thin ice layers, the wavefield has two frequency regimes: Most of the phases specifically related to the ice layer (Crary phase, longitudinal and flexural waves) have a maximum frequency between 0.1 and 0.5 Hz (for thicknesses of 20 and 5~km resp.), depending on ice thickness. Body waves from the mantle are not limited in maximum frequency and should thus be clearly distinguishable (see the spectrogram in fig. \ref{fig:spectrogram}). Scattering within the ice layer enhances this effect by damping high frequency ice body waves, with long paths in the ice layer, generally ice-body waves fade away after a short distance.
\begin{table}
\begin{tabular}{l|ccccc}
Wave type & frequencies & polarization & symmetry & group vel. $U$ & phase vel. $c$\\ \hline
Love wave & high & transversal & antisymmetric & $v_{\mathrm{S}} \sqrt{1-\left(\frac{n v_{\mathrm{S}}}{2fd}\right)^2}$ & $v_{\mathrm{S}} / \sqrt{1-\left(\frac{n v_{\mathrm{S}}}{2fd}\right)^2}$\\
Toroidal mode & very low & transversal & symmetric & $\approx v_{\mathrm{S}}$ & $\approx v_{\mathrm{S}}$\\
Crary wave & monochromatic & radial & n/a & $\approx v_{\mathrm{P}}\sqrt{1 - v_{\mathrm{S}}^2/v_{\mathrm{P}}^2}$ & $v_{\mathrm{P}}$\\
Longitudinal & low & radial/vertical & symmetric 
& $\approx2 v_{\mathrm{S}}\sqrt{1 - v_{\mathrm{S}}^2/v_{\mathrm{P}}^2}$ & $2 v_{\mathrm{S}}\sqrt{1 - v_{\mathrm{S}}^2/v_{\mathrm{P}}^2}$\\
Flexural wave & low & vertical & antisymmetric 
& $c - \lambda \frac{\mathrm{d}c}{\mathrm{d}k}$& $ v_{\mathrm{S}} \sqrt{\frac{8\rho_{\mathrm{ice}}(kd)^3}{3\rho_{\mathrm{water}}}  \frac{1 - v_{\mathrm{S}}^2/v_{\mathrm{P}}^2}{1+2kd\rho_{\mathrm{ice}}/\rho_{\mathrm{water}}}}$
\\
Rayleigh wave & high & radial/vertical & n/a & $0.9194 v_{\mathrm{S}}$ & $0.9194 v_{\mathrm{S}}$ 
\end{tabular}
\caption{Wave types encountered in ice floating on water. $v_{\mathrm{P}}, v_{\mathrm{S}}$ are the wave velocities in ice, which are assumed to be constant over the depth. $\rho_{\mathrm{ice}}, \rho_{\mathrm{water}}$ are the densities of ice and water The definition of symmetry is analog to Lamb waves, i.e. symmetry with respect the the center axis of the ice layer. }
\label{tab:wave_types}
\end{table}

\subsubsection{Water phases}
Due to the large velocity contrast between compressional wave velocity in the ice and in the water, no phases like PFP exist in shallow ocean worlds like Europa, while they would be a dominant signal in the late seismogram on Ganymede, Titan. On Enceladus, they will be present, if its ice is thin enough. Attenuation in water is very small, so phases that are multiply reflected at the top and bottom of the water layer (Pf, Pf2...) are observable almost over the whole distance range. Their polarization is mainly on the radial component.

\subsubsection{Crustal and Mantle phases}
Body waves in the rocky interior cover the whole frequency range. On a thin-ice world like Europa, they dominate the seismic signal above 0.1 Hz. Because of this, a broadband seismometer, sensitive enough to measure in the whole frequency range, can estimate the wave type from its frequency content alone. It also means that, while displacement amplitudes are dominated by ice phases after their arrival, body waves might still be recovered from the high-frequency content of the seismogram.

The Scholte wave is one particular crustal phase occurring at the ocean bottom. This wave shares most characteristics of a Rayleigh wave, but is defined on a solid-fluid interface \citep{Scholte1947}. Its phase velocity is frequency-dependent, but below the shear velocity in the solid and the sound velocity in the fluid, whatever is lower \citep{Rauch1980}, i.e. below 1.8~km/s. It has been used in marine exploration on Earth to estimate the shear wave velocity of the uppermost layers below the sea floor \citep{Nolet1996, Kugler2005}. If it can be detected, a detailed dispersion analysis might be used to constrain the existence of hydrated low-velocity layers on the sea floor. As it can be seen in Figure \ref{fig:global_stack}, it shows strong water multiples.

\subsubsection{Distance regimes}
In terrestrial seismology, several distance regimes are defined, because seismograms change their first order shape considerably from one to the next \citep{KennettSeismicWavefieldII}:
\begin{description}
\item[Local (0 to 100~km)] For shallow events, the ground displacement is controlled by the near field of the source and significant permanent deformation occurs at the surface. The local regime can be considerably larger for mega-thrust events \citep{Grapenthin2011}.
\item[Regional (100--1000~km)]
Crustal body wave phases arrive first. Crustal S-waves and surface waves form one wave train (Lg).
\item[Far regional (1000--3000~km)] Mantle body wave phases arrive first and are separated from the surface wave train, but due to the upper mantle discontinuities, multiple (triplicated) phases arrive within a short time window.
\item[Teleseismic (3000--9000~km)] Mantle body wave phases are clearly separated from each other and are visible in the seismogram as well-defined sharp pulses. Direct body waves do not sense the core.
\item[Core shadow (beyond 9000~km)] Direct P-waves are shadowed by the core. Multiple complex core phases arrive first, followed by surface-reflections (PP).
\end{description}
To summarize, the distance regimes define whether the first arriving phases have mainly travelled through the crust, the complex upper mantle, the relatively homogeneous lower mantle or the core. The situation on icy ocean worlds is similar. We propose to separate the following regimes:
\begin{description}
\item[Local] Near field effects are dominant, including permanent displacement.
\item[Regional] Body waves that traveled through the ice layer only arrive first.
\item[Ocean shadow] Direct body waves are masked by the curvature of the planet's surface. The first-
arriving phases are longitudinal and Crary waves and Love waves.
\item[Teleseismic range] Mantle body waves arrive first
\item[Core shadow] The planetary body's core masks direct body waves.
\end{description}
The distance range of these regimes depends mostly on the thickness of the ice layer. An exception is the local regime, whose size depends mainly on the magnitude of the event, but also on the local velocity structure. On thin-ice worlds like Europa, considerable coseismic displacement will occur over hundreds of kilometers, but its long periods will not be detectable with a space-ready seismometer. Therefore, we do not discuss this region in detail here.

The distance regimes for a set of icy ocean worlds can be found in Table \ref{tab:distance_regimes}. 
For the thin-ice worlds like Europa, it is unlikely that an event will randomly fall into the regional area around a lander, where direct ice phases are well-recordable (unless a specifically active region has been chosen for the lander). Roughly 5\% of randomly distributed events will be in the ocean shadow range, where special ice layer phases will be the first arrivals. The remaining events will be beyond that in the range, where mantle and core body waves arrive first.
\begin{table}
  \begin{tabular}{lcccc}
  Moon &  & regional & ocean shadow                   & teleseismic \\ \hline
  Europa & 5~km ice & $0-5^\circ$ (0.2\%) & $5 - 30^\circ$ (6.5\%)& $30 - 150^\circ$ (86.6\%) \\
  Titan & 33~km ice & $0-20^\circ$ (3\%) & $20-55^\circ$ (18.3\%) & $55 - 160^\circ$ (75.7\%) \\
    & 114~km ice & $0-50^\circ$ (17.9\%) & - (0\%) & $50 - 150^\circ$ (75.4\%) \\
  Ganymede 104~km & & $0-60^\circ$ (25\%) & - (0\%) & $60 - 120^\circ$ (50\%) \\
  Enceladus 15~km &  & & $5-80^\circ$ (41.1\%) & $80-180^\circ$ (58.7\%)\\ \hline
  Earth & & $1-10^\circ$ (0.75\%) & $10-30^\circ$ (5.9\%) & $30-105^\circ$ (56.4\%)\\
  \end{tabular}
  \caption{Distance ranges of seismograms regimes on different icy worlds. Epicentral distances between earthquake and receiver are noted in degrees. The number in parentheses notes the fraction of the total planetary surface that is covered by this distance regime. If we assume a random distribution of events on the surface, this is also the probability that any given event will fall into this distance range from a lander.
  For comparison, the distance ranges on Earth are also noted, where "ocean shadow" corresponds to "far regional earthquakes".}
  \label{tab:distance_regimes}
\end{table}

\subsection{Single-station seismology on icy moons}
\label{sec:single}
The first efforts at seismology on icy moons will probably be conducted from a single station. Single-station seismology was common on Earth for decades, and many fundamental discoveries, like the existence of the inner core, were derived from observations of earthquakes on single observatories \citep{Lehmann1936}. Other fundamental work studied one earthquake at multiple receivers, which is conceptually similar \citep[e.g. the discoveries of the core and the Mohorovicic discontinuity by][]{Oldham1906, Mohorovicic1909}. It has long been known that the parameters of earthquakes can be determined from one station alone \citep{Ekstrom1986, Wu2006}. A large body of work on modern single-station seismology has been done in the context of the Mars seismometer on the InSight lander \citep{Panning2015, Khan2016, Bose2016}. The location of a marsquake is determined by the polarization of P and surface waves and the time difference between specific phases.

\begin{figure}
  \centering
  \includegraphics[width=0.9\textwidth]{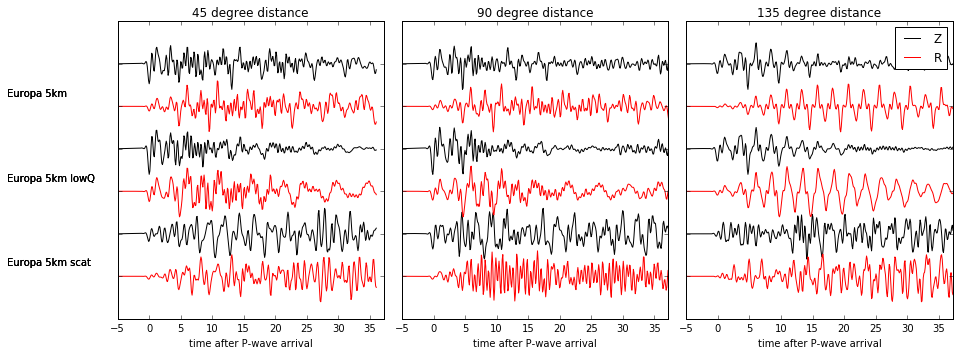}
  \caption{Coda of a Pn wave measured at 45, 90 and 135 degree distances for various Europa models. Black lines correspond to vertical displacement (Z-component), red to horizontal pointing away from the source (R-component).
 Clearly visible are the reverberations from the ice layer, which will serve to estimate the ice thickness from a single seismogram.}
  \label{fig:Pcoda}
\end{figure}

\begin{figure}
  \centering
  \includegraphics[width=0.49\textwidth]{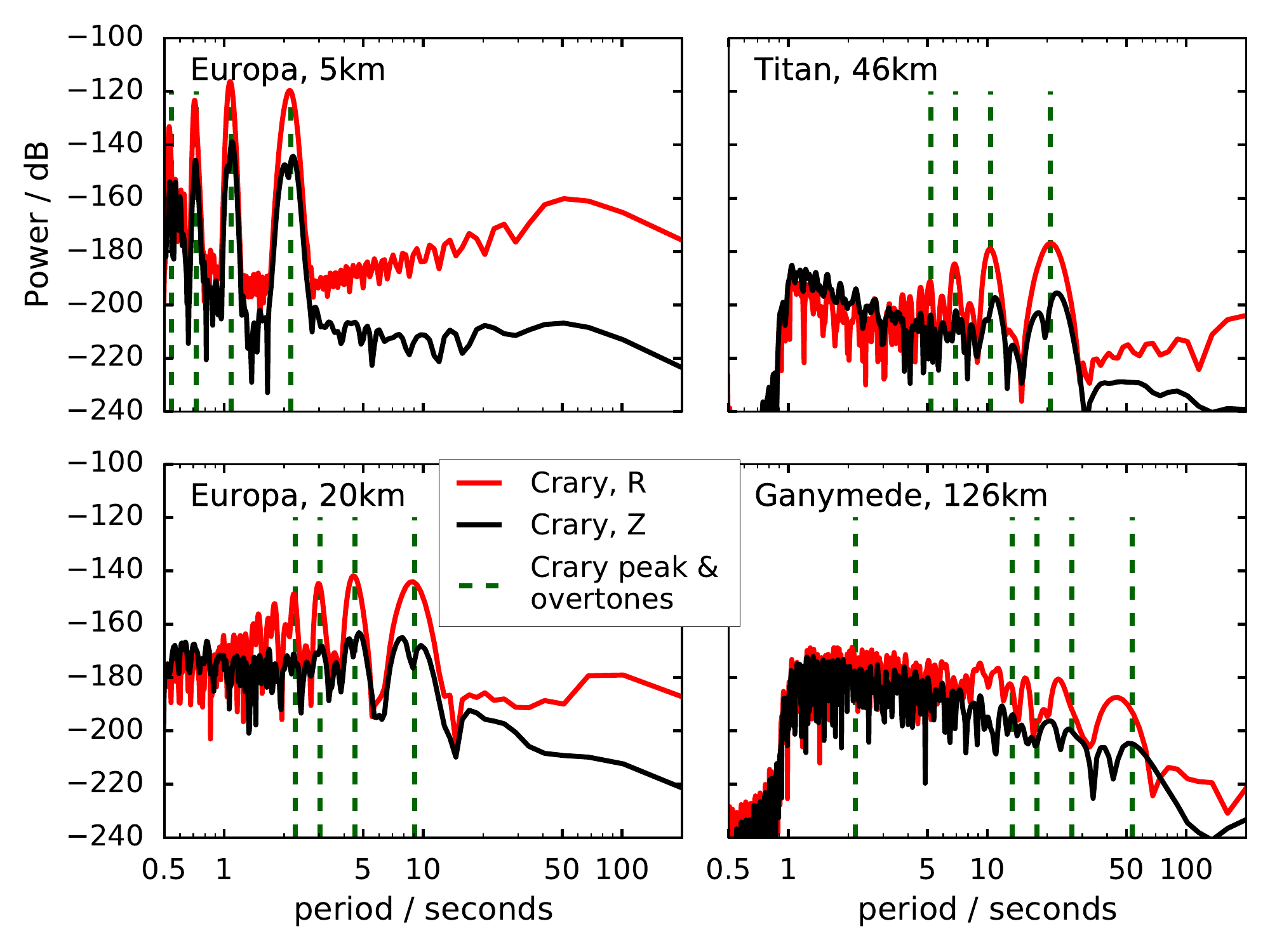}
  \caption{Spectra of a time window around the arrival of the Crary phase for two Europa models of ice thicknesses of 5 and 20~km and Titan and Ganymede. The period of the main mode and overtones of the Crary phase are proportional to the ice thickness (eq. \ref{eq:Crary_frequency}).}
  \label{fig:Crary_spectra}
\end{figure}

Because the longitudinal and the Crary wave are prominent features in the seismogram of Europa, they can be used for fast determination of the backazimuth of the event (the direction as seen from the seismometer; see appendix \ref{app:baz} for a worked example). Their energy loss due to attenuation is relatively low, so they propagate around the planet multiple times. From the time difference between these orbits, the distance can be determined. On thick-ice worlds, the Rayleigh wave can take this role.
This constraint is best achieved using distant events, using the following work flow:
\begin{enumerate}
\item Estimate the ice thickness from converted waves in the coda of body waves of distant events. This analysis needs no event localization and not even a clear identification of phases. Figure \ref{fig:Pcoda} shows the coda of Pn waves in different Europa models. Its coda contains clearly recognizable multiples, whose time difference is the ice thickness divided by the wave velocity. 
The effect is clearest for very large distances (135 degree), where the incident angle is almost vertical. The horizontal component seismograms have their maximum energy several seconds after the vertical seismogram. The two lower models in fig. \ref{fig:Pcoda} have been modified to simulate the effect of a highly attenuative ice layer (lowQ) or of a low-attenuation, high-scattering model, similar to the lunar crust (scat). Here, the reverberations become much less pronounced, but are still visible, especially on the long-distance horizontal components. This method will work best for large distances, where Pn is a clear arrival. The 45 degree seismograms contain triplicated phases, which creates the very complex waveform.

\item With the constraint from step 1, refine the thickness estimate from the characteristic harmonic frequency of the Crary wave (for ice layers < 50~km) and estimate P- and S-wave velocities in the ice. Figure \ref{fig:Crary_spectra} shows spectra of the Crary wave from simulated waveforms for two Europa models, Titan and Ganymede. A time window after the high amplitude longitudinal wave LL was selected, where the Crary wave can easily be found as a harmonic signal, even without identifying any other phases. For all realistic ice thicknesses of Europa, the Crary peaks are very prominent in the spectra of the R-component and can be used to determine the ice thickness reliably. For ice layers of more than 100 km, the peaks are less prominent and the implicit assumption of a thin ice layer with constant wave speeds is not valid anymore, as seen for the Ganymede model.

\item With an ice layer model, locate local events, using the time difference between Pe and Se phases, as well as reflections from the ice-ocean boundary (PeS or PeP), once the ice thickness has been constrained. The distance of a local event with P and S arrival times $t_P, t_S$ is
\begin{equation}
x = \frac{t_S - t_P}{1/v_S - 1/v_P}\approx (t_S - t_P)\cdot 4~\mathrm{km/s}.
\end{equation}

\item From multiple-orbit Crary, longitudinal or Rayleigh waves, locate distant events. Appendix \ref{app:baz} shows an example of determining the direction of an Europaquake using the polarization of body waves and the energy content of the horizontal channels. If minor arc ($t_1$), major arc ($t_2$), and single orbit ($t_3$) arrival times of any surface wave are known, the distance of the event is \citep{Panning2015}
\begin{equation}
\Delta = 180^\circ \left( 1- \frac{t_2-t_1}{t_3-t_1} \right).
\end{equation}
Note that the group velocity of the wave is not needed. If it is known, only two of the arrival times are needed. 

\item Identify ocean multiples in the coda of body waves of distant events and constrain the ocean depth by estimating the two-way travel time of reflected phases. 

\item Identify Scholte waves at the ocean bottom to constrain the uppermost crustal structure below the sea floor from spectral peaks or dispersion curves (see Sect. \ref{sec:Titan} for an example on Titan).

\item Identify as many body wave arrivals as possible to constrain overall mantle velocity structure with the event locations from step 4, analogous to the method proposed by \citet{Khan2016} for Mars.

\item If direct ocean phases (PFP, SFS) have been identified together with other body waves, the speed of sound in the ocean can be estimated, constraining the ocean chemistry.
\end{enumerate}
The process sketched here will be an iterative one in practice. Steps 1 and 2 can be done even in noisy seismograms without clearly identified phases or located events. Steps 3 and 4 need some identified phases and benefit from first estimates on ice thickness and ice velocities. The steps 5-8 need first estimates of event distances to work, but their results can be used to better constrain the velocity models in turn.
While the process is described here only for a single event, observations of multiple events will be necessary to obtain reliable results. Recent estimates of the seismicity induced by tidal cracking on Europa \citep{Panning2017} suggest at least one globally observable event per week. The seismicity on other ocean worlds is not yet estimated.


\section{Icy ocean worlds in the solar system}

\subsection{Europa}
\label{sec:Europa}
Europa has been the subject of the most detailed studies to date, and a lander mission could take place in the coming decades \citep{Hand2017}. Therefore, we discuss wave propagation and measurability of seismic waves in detail here.

\subsubsection{Overview}
Compared to its radius, Europa has a thinner ice layer than all other icy ocean worlds. This creates pronounced ice-layer-specific seismic phases. The transition from a ice-Rayleigh wave to flexural waves occurs between 0.1 Hz (for 20~km ice thickness) and 0.5 Hz (for 5~km ice) \citep{Panning2006}, which would be well observable with a wide-band seismometer, as the SP instrument used for InSight \citep{Pike2016} and proposed as the baseline instrument for the NASA Europa lander \citep{Hand2017}.

Apart from the thickness of the ice layer, which can be measured as described above, or from the Crary phase, key science questions on Europa are the depth and composition of the ocean, the presence of liquid or mushy pockets in the ice and the potential of material transfer between the rocky interior, the ocean and the ice layer. The depth of the ocean can be determined well from the time delay between ocean multiples (PfPn, Pf2Pn, ...) in the P-wave coda of teleseismic Europaquakes. The ocean chemistry can be estimated from the sound speed in the ocean, which is increased by higher salt contents. This increases the curvature of the SFS path, resulting in higher amplitudes for SFS at short periods; a phase that would not be observable in a pure water ocean. This could be used to constrain the ocean chemistry, once the ocean depth and thickness have been determined.
Liquid pockets close to the lander would result in scattering of seismic energy from certain incidence angles. Motion of liquids in the ice shell would create seismic signals similar to geysers on Earth \citep{Kedar1998}. The sharpness of the ocean-rock interface, including potential hydrated layers could be estimated from Scholte waves on the seafloor.

Due to Europa's modest size, multiple planetary orbits of seismic ice-phases, especially the longitudinal waves, would be observable with an SP-like instrument, allowing for a straightforward inversion for the distance of a seismic event. Due to the thin ice layer, no direct SH waves are possible beyond 100 km, so that the transverse component will be very distinct from the radial component, allowing an estimation of the event backazimuth. 

In a companion paper \citep{Panning2017}, the background seismicity from tidal cracking is estimated, showing that it is detectable with a seismometer. This seismicity will create a significant seismic hum, which exceeds noise floor of industry standard Trillium compact instruments \citep{Ringler2010} at periods between 1 and 10 seconds. It could be be used for ambient noise analysis techniques, to constrain the thickness of the ice layer or mantle discontinuities. 

Another important seismic source will be surface tectonics. The young surface of Europa shows clear signals of subduction \cite{Kattenhorn2014}, collapses of cavities \citep{Walker2015} and potentially plume activity, which would all have specific seismic signatures \citep{Vance2017b}. While no atmosphere exists to create microseismic noise, the vicious circulation in the ocean \citep{Soderlund2013} might just be strong enough to create a detectable long-period signal \citep{Panning2017}. 
\subsubsection{Measurability}
\begin{figure}[ht]
\includegraphics[width=0.9\textwidth]{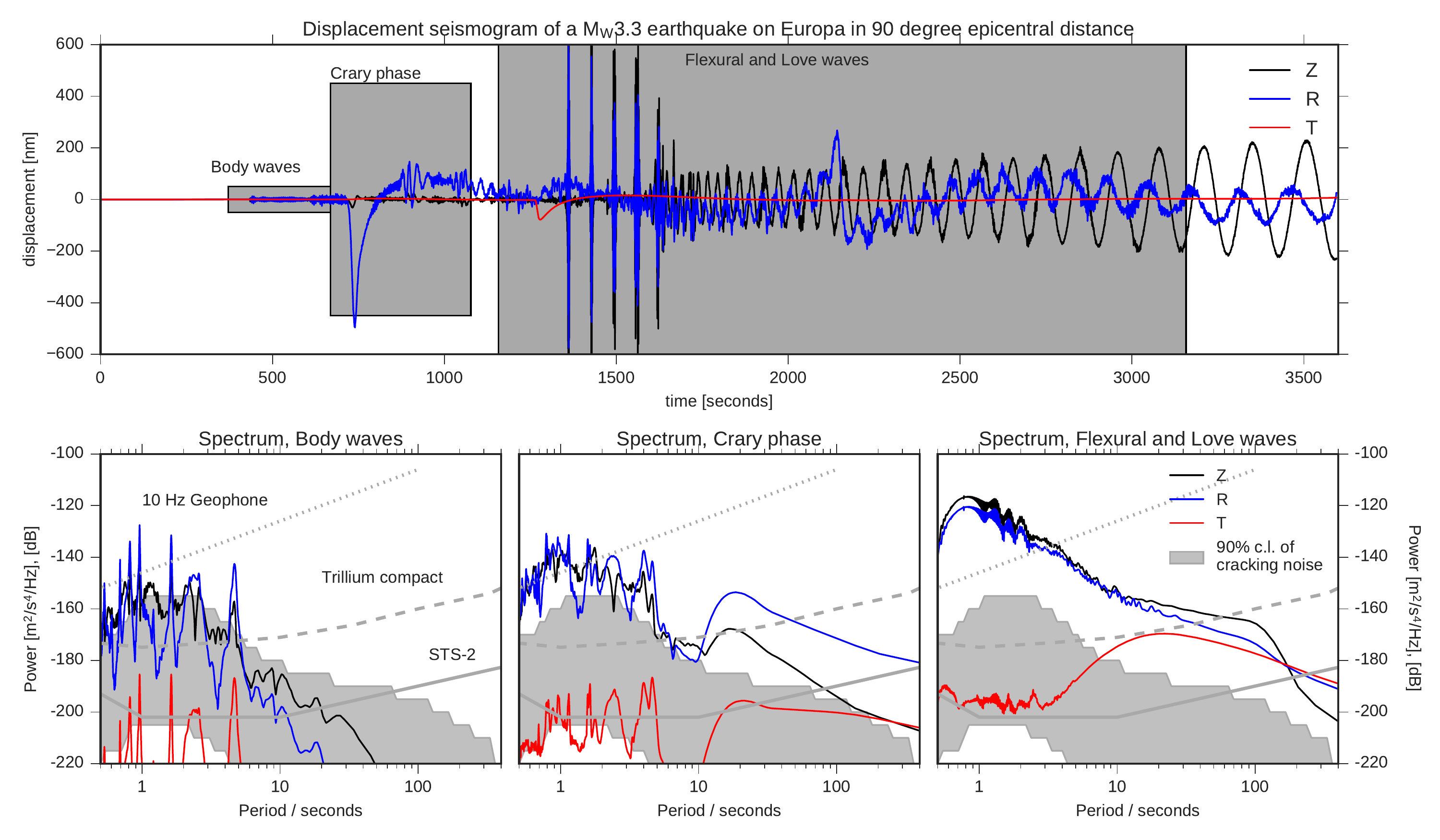}
\caption{Measurability of seismic phases on Europa: The upper plot shows an example seismogram of a $M_W$ 3.3 event in 2 km depth in the ice shell in 90 degree distance. The velocity model has a 5~km thick ice layer, with high Q (low attenuation) and strong scattering (K\'arm\'an medium, 5~km correlation length, amplitude 10\%). Three significant time windows are selected: A) a body wave window between 380 and 690 seconds, which contains only waves that traveled through the mantle. B) a window with the longitudinal wave, a non-dispersive wave within the ice layer that consists of pure vertical particle motion and C) a window in the later seismogram, where surface-wave-like phases arrive on all components. The lower plots show spectra of these three time windows vs the established self-noise models of typical instruments, based on \cite{Ringler2010}. To account for effects of the radiation pattern and source depth, the spectra are averages of 100 events with depths between 0 and 5 km, with random focal mechanisms. The grey area in the spectra corresponds to the 1st and 9th decile of the noise model C in \citet{Panning2017}, simulating seismic noise due to tidal cracking.}
\label{fig:measurability}
\end{figure}
Since a seismometer on Europa will probably have characteristics not exceeding those of a Trillium compact instrument, we tested which seismic phases could be detected for a magnitude 3.3 event at a distance of 90 degree (fig. \ref{fig:measurability}). Body waves exceed the instrument self-noise by 20~dB between 0.1 and 2 Hz, and will therefore be observable. The longitudinal phase, which is crucial for event location, exceeds the self-noise by 30~dB. The late seismogram on R and Z, including Rayleigh and flexural waves, can be measured for periods up to 50s. The Love wave on T, which is confined to the ice layer and therefore strongly affected by attenuation and scattering in the ice, is not observable for most focal mechanisms. Based on the seismicity models in \citet{Panning2017}, a magnitude 3.3 event can be expected to occur somewhere between once per week and once per month. Compared to Mars or Earth, an instrument on Europa will not be affected by atmospheric noise, so instrument self-noise is the reference to measure signal strength against.
\subsection{Titan}
\label{sec:Titan}
\begin{table} \center
\caption{Titan model used for detectability test of an IceVI layer as shown in fig. \ref{fig:Titan_iceVI}}
\begin{tabular}{lcccc}
Layer & thickness [km] & v$_P$ [km/s] &  v$_S$ [km/s] & $\rho$ [kg/m$^3$] \\ \hline
Iceh & 46 & 3.9 & 2 & 925\\
ocean & 410 & 1.55 -- 2.55 & -- & 1020 -- 1170\\
IceVI & 5.5 & 4.4 & 2.26 & 1394\\
mantle &  & 7.99 & 4.54 & 3526 
\end{tabular}
\label{tab:Titan_iceVI}
\end{table}
\begin{figure}
\includegraphics[width=0.9\textwidth]{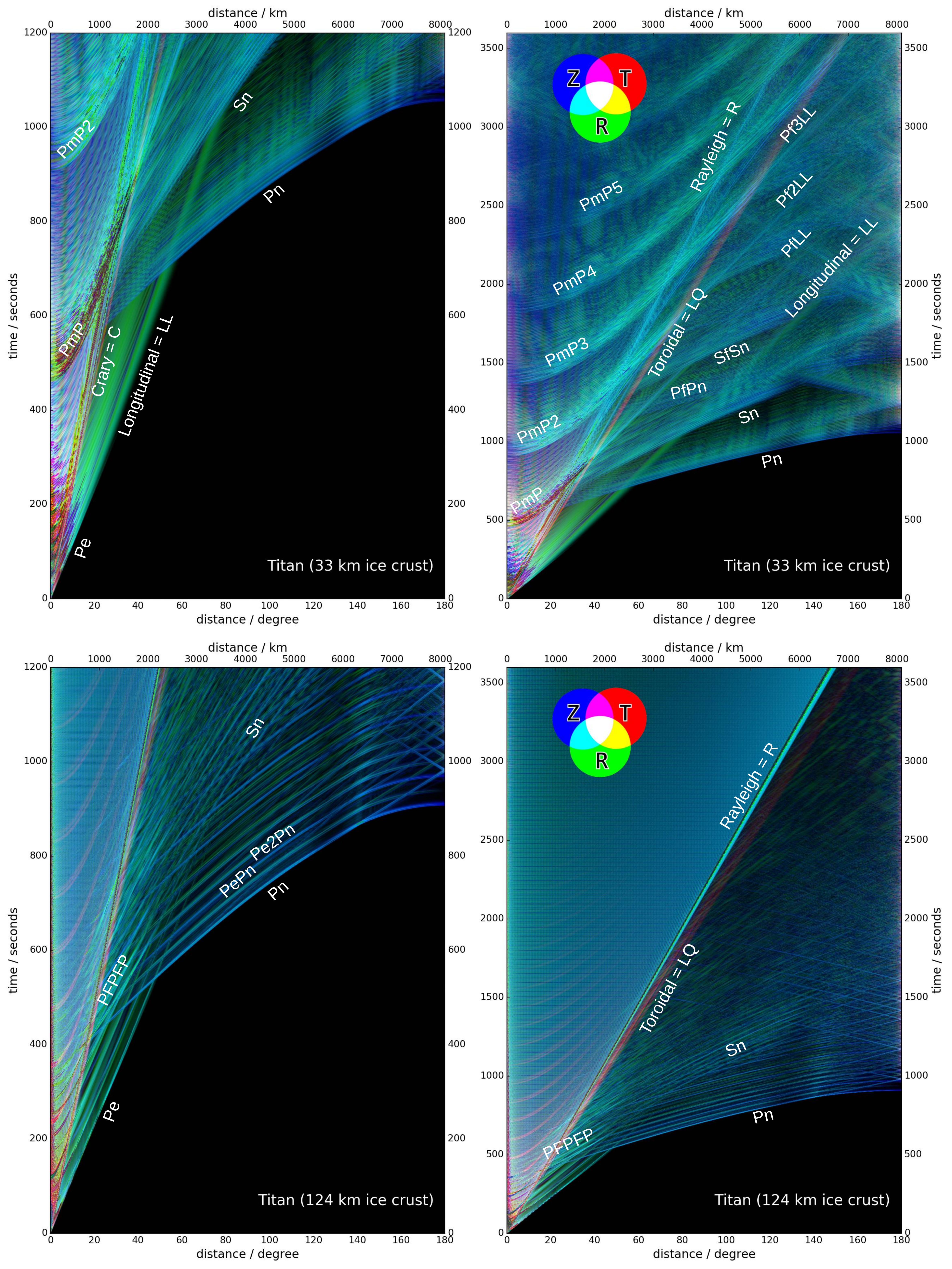}
\caption{Global stack of seismograms for a Titan model of 46 km ice thickness (pure water ocean and 270K in table 7 of \citet{Vance2017}, top, compared to one with 119 km ice (bottom, model with 3\%NH$_3$ ocean and 255K in table 7 of \citet{Vance2017}). The thinner ice model has a seismic wavefield resembling that of Europa, with a dominant longitudinal phase that is the first arriving phase between 20 and 55 degrees. In the thick-ice model, seismic phases from high-pressure ice are negligible, and the wavefield is more similar to a terrestrial planet, where body waves always arrive first. In contrast to Earth or Mars, every phase is followed by a large number of multiples from ice and/or the ocean.}
\label{fig:Titan_stack}
\end{figure}
\begin{figure}
\includegraphics[width=0.9\textwidth]{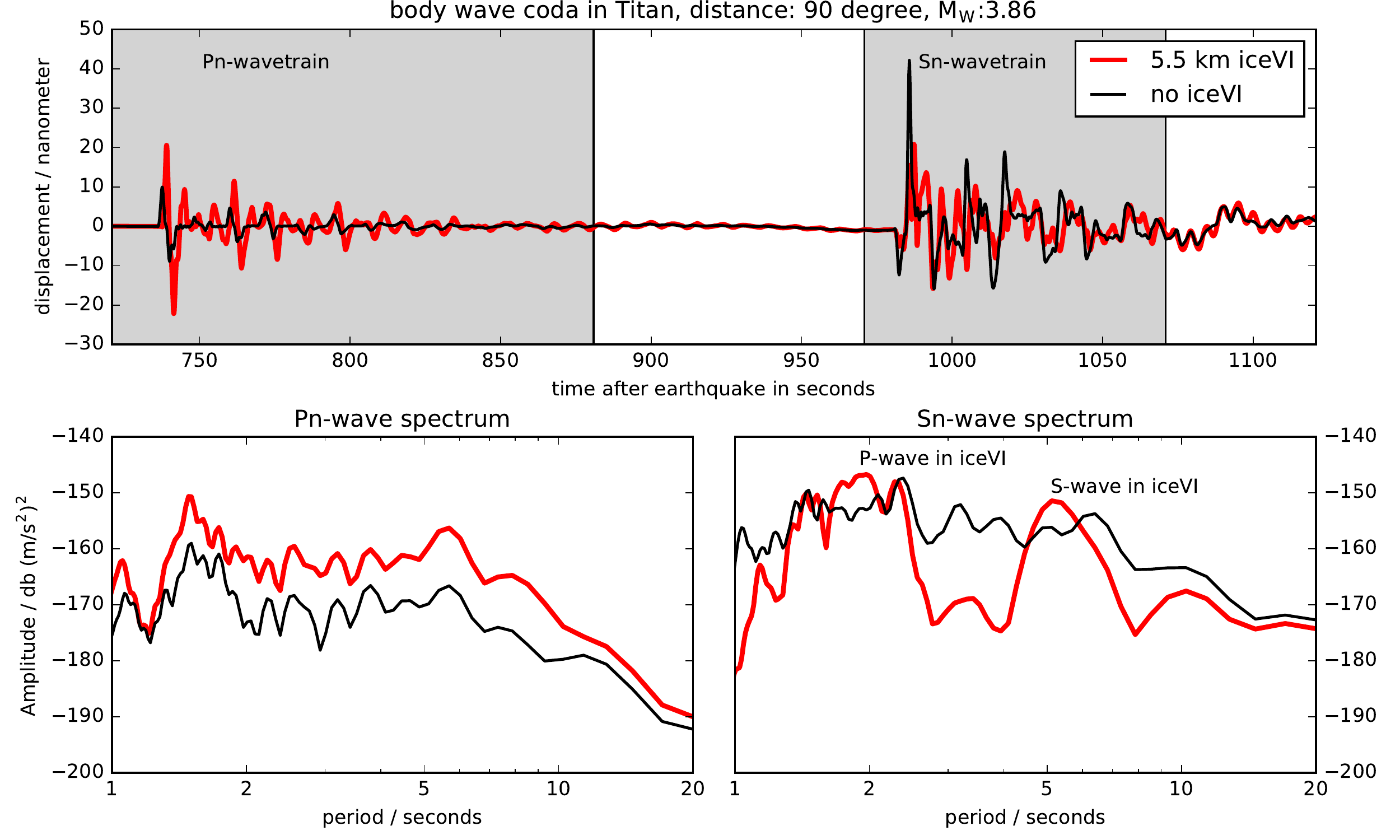}
\caption{Detectability of an high-pressure ice layer (IceVI) below the ocean from body and Scholte waves. The upper plot shows the vertical displacement of an M3.8 event in 90 degree distance and 5 km depth for two interior models with 46~km ice thickness. The red lines correspond to a model with a 5.5~km IceVI layer the bottom of the ocean. The model displayed with black lines is identical, but the mantle starts directly below the ocean. The peaks in the S-spectrum correspond to reverberations in the ice layer and are clearly identifiable.}
\label{fig:Titan_iceVI}
\end{figure}
Titan is predicted to have water and ice layers with total thickness around 480~km above a low-density silicate mantle. The ice covering is interpreted to be 55-80~km based on the observed Schumann resonance \citep{beghin2012analytic}. Following \citet{Vance2017}, we consider thicknesses in a slightly wider range between 46 and 118~km. Depending on the ocean's temperature profile with depth, high-pressure ice phases might be present below the ocean, with a total thickness of up to 230~km or not present at all. Depending on the ocean temperature, the ice thickness may be so high that the longitudinal, Crary, and flexural waves are restricted to very long periods, unobservable with a realistic instrument. Because the rocky mantle makes up more than 80\% of the diameter in all models, the teleseismic range dominates on Titan (see fig. \ref{fig:Titan_stack}). 

An important question for habitability in Titan is the existence of high-pressure ice phases at the bottom of the ocean. If they are present, they may form a barrier to water-rock interactions that regulate redox and ocean composition, and which are thereby critical for supporting life. While intra-ice convection might be able to transfer ions \citep{Choblet2017}, the absence of a high-pressure ice layer could benefit habitability by providing a direct water-rock interface, potentially with life-supporting hydrothermal systems analogous to those on Earth's seafloor. Because such a high-pressure ice layer is not detectable with radar or gravitational moment-of-inertia measurements, seismological measurements provide an unique probe. 
As was argued in \citet{Vance2017}, the thickness of surface ice above the ocean and high-pressure ice below are strongly correlated and both mainly controlled by the ocean ion content and temperature profile. According to these simulations, every model with a surface ice thickness of more than 50~km must also contain high-pressure ice layers. Seismic methods could constrain surface ice and high-pressure ice thicknesses independently, and thereby constrain ocean chemistry and temperature. 

As one example, Figure \ref{fig:Titan_iceVI} shows a comparison of waveforms for two thin surface ice models (46~km) on top of a 410~km deep ocean with 3.3 weight percent NH$_3$ (similar to the 3 wt percent NH$_3$ model with 264~K in table 8 of \citep{Vance2017}). For this surface ice thickness, high pressure ice can neither be confirmed nor excluded. The figure compares seismograms of this model with one where the iceVI layer has been replaced by mantle (see table \ref{tab:Titan_iceVI}).  Because there is no distance where high-pressure ice phases arrive first, their presence will have to be inferred from the coda of mantle body waves. The strong velocity and density increase at the ocean-mantle interface (assuming that no iceVI layer is present) causes a strong impedance contrast, which is weakened if there is an intermediating layer. The lower two plots in fig. \ref{fig:Titan_iceVI} show acceleration spectra of a window around the P-wave arrival (left gray patch in upper subfigure) and the S-wave arrival (right gray patch in upper subfigure). The amplitude of the Pn wave train in a model with an ice layer is two times higher than in the model without the layer. Since it will be difficult to reliably determine the magnitude and focal mechanism of a seismic event with just one station, this effect may go unnoticed. The spectrum of the S-waves (lower right) however, shows clear effects of the iceVI layer: While the spectrum of the simple model is almost flat between 1 and 10 seconds, the iceVI layer produces two clear peaks at 2 and 5.1 seconds, corresponding to reverberations of P and SV waves. This signal should be easily distinguishable, even without a detailed source model.

Titan is the only icy ocean world with an atmosphere. Due to the atmosphere's low density in comparison with the solid and liquid portions, the direct effect on the seismic wave propagation in the planet will be minimal; however, it could serve as a source for seismic signals. Wind waves on Titan’s methane lakes have been predicted to reach significant wave heights of 0.2 m at periods of 4 seconds \citep{Lorenz2012} for winds of 1m/s.
 The handful of Cassini measurements available through spring did not show evidence of waves above a few millimeters in height e.g. \citep{Hofgartner2014, Zebker2014} but these
may simply have been on calm days. If or when 0.2m waves form, they create pressure on the sea floor of the order of 100~Pa, which is comparable to pressure variations on the abyssal sea floor on Earth due to ocean wave interference \citep{Cox1984, Davy2014, Staehler2016}. Because the latter creates a globally observable seismic signal \citep{Kedar2008}, it must be assumed that measurable ocean-generated microseisms are possible on Titan. Surface waves from these signals could be used to constrain the uppermost layers below the surface. Note, that while Titan's seas occupy 1\% of Titan's surface area total, they cover
some 12\% of the terrain northward of 55 degree N \citep{Hayes2016}, so microseisms might be expected to be a much more prominent feature of the seismic 
environment in Titan's high northern latitudes than elsewhere, and more so during the 
summer when winds are expected to be strongest.

Pressure changes in the atmosphere might also create seismic signals directly, as observed on long periods on Earth \citep{Peterson1993, Beauduin1996}. Local phenomena, such as "dust devils" on Earth and potentially Mars \citep{Lorenz2015a} might also be observable. Seismic waves may also create atmospheric signals that might be detectable from orbit, as observed on Earth and proposed for Venus \citep{Lognonne2016}.
\subsection{Ganymede}
\begin{figure}
\includegraphics[width=0.95\textwidth]{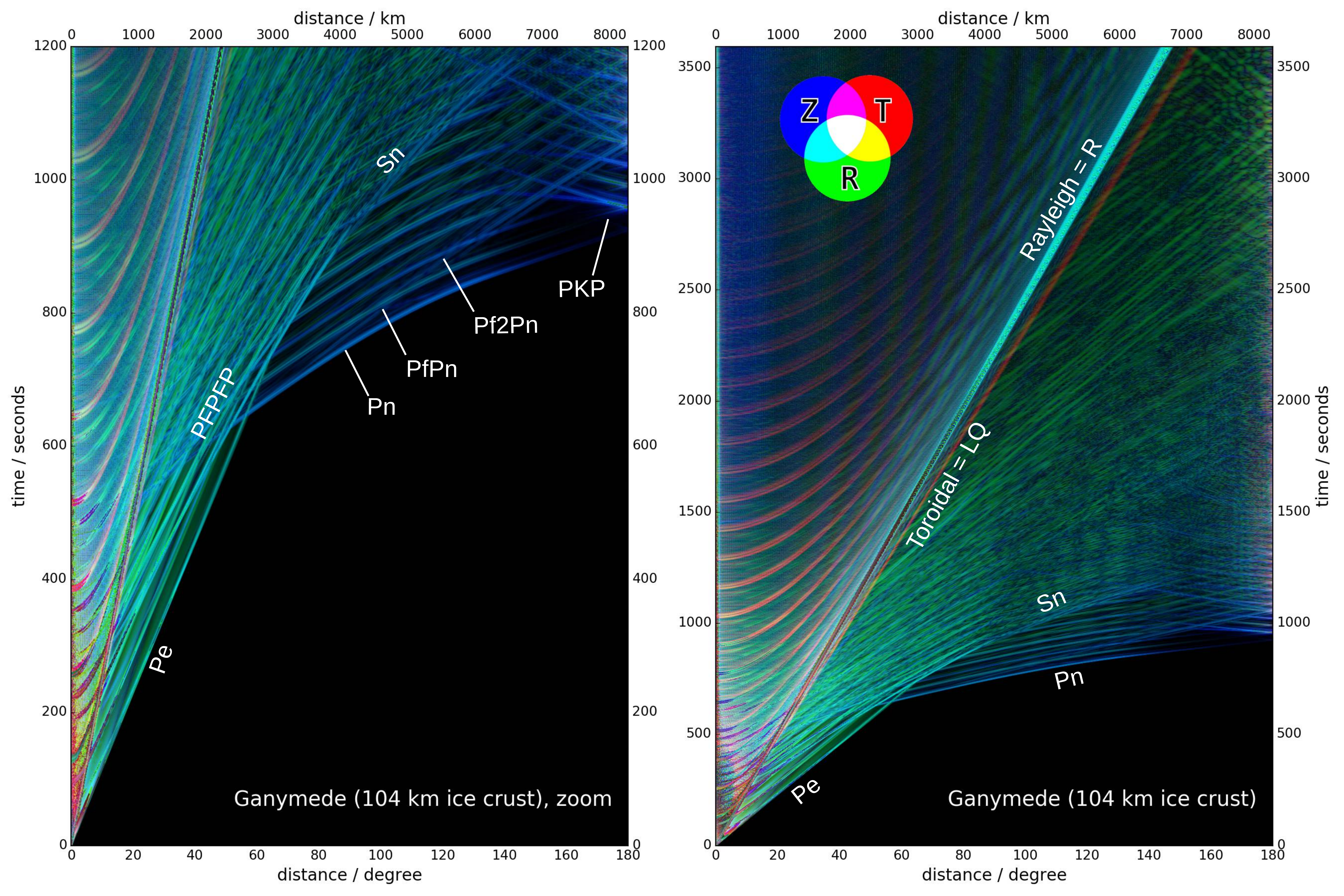}
\caption{Global stack of seismograms for a Ganymede model of 104~km ice thickness. The thick ice layer creates nicely separated body wave reverberations in the ice. Seismic waves that bottomed in the high pressure ice layer arrive in the coda between 20 and 90 degree distance. They seem to connect to Sn at 90 degree, since the P-velocity in the high-pressure ice is very close to the S-velocity in the mantle. The overall wavefield is very similar to a cold-ocean Titan.}
\label{fig:stack_Ganymede}
\end{figure}
The density structure inferred from moment of inertia measurements suggests that Ganymede has water layers in different forms, with total thickness of about 800~km above a rocky mantle and an iron core \citep{Vance2014}. The uppermost ice layer is assumed to have a thickness of more than 50~km, floating on a liquid ocean of more than 100~km depth. Due to the high gravity and abundance of water, pressure at the ocean floor creates layers of ice V and VI 10s to 100s of km thick, between the liquid ocean and the rocky mantle. Similar to Titan, the existence and thickness of these layers can only be measured by seismology. Since both are controlled by ocean chemistry and temperature, this creates a direct geophysical observable for habitability conditions of the ocean \citep{Vance2017}.

The thick ice layer means that longitudinal and Crary phases are relatively weak and constrained to long periods, which makes them unobservable without a true high-sensitivity broadband seismometer. The ice thickness therefore would have to be constrained from multiples in the coda of body waves. Compared to Titan, the thickness of the combined water layers is 30\% greater, which clears the coda of body waves. As fig. \ref{fig:stack_Ganymede} shows, the wavefield is nevertheless dominated by water multiples. The analysis presented in the previous section to detect high-pressure ice phases could be done in a similar fashion for Ganymede, though the high-pressure ice layers are much thicker and more heterogeneous than on Titan which will shift the resonances to longer periods.

Ganymede's interior is likely to be at least as seismically active as that of Earth's moon, where continuous seismicity was observed by Apollo instruments \citep{Vance2017}. A major question at Ganymede involves the presence of a liquid iron core, which seems to be required by Galileo observations of an intrinsic dipolar magnetic field \citep{Kivelson2002}. Melt could be present in the overlying rock as well. This study necessarily focuses on the initial problem of identifying radial boundaries closer to the surface, but a study focused on the goal of long-term exploration of Ganymede would need to evaluate the ability to probe the deeper interior to understand Ganymede's composition and current thermal state.
\subsection{Enceladus}
\begin{figure}
\includegraphics[width=0.95\textwidth]{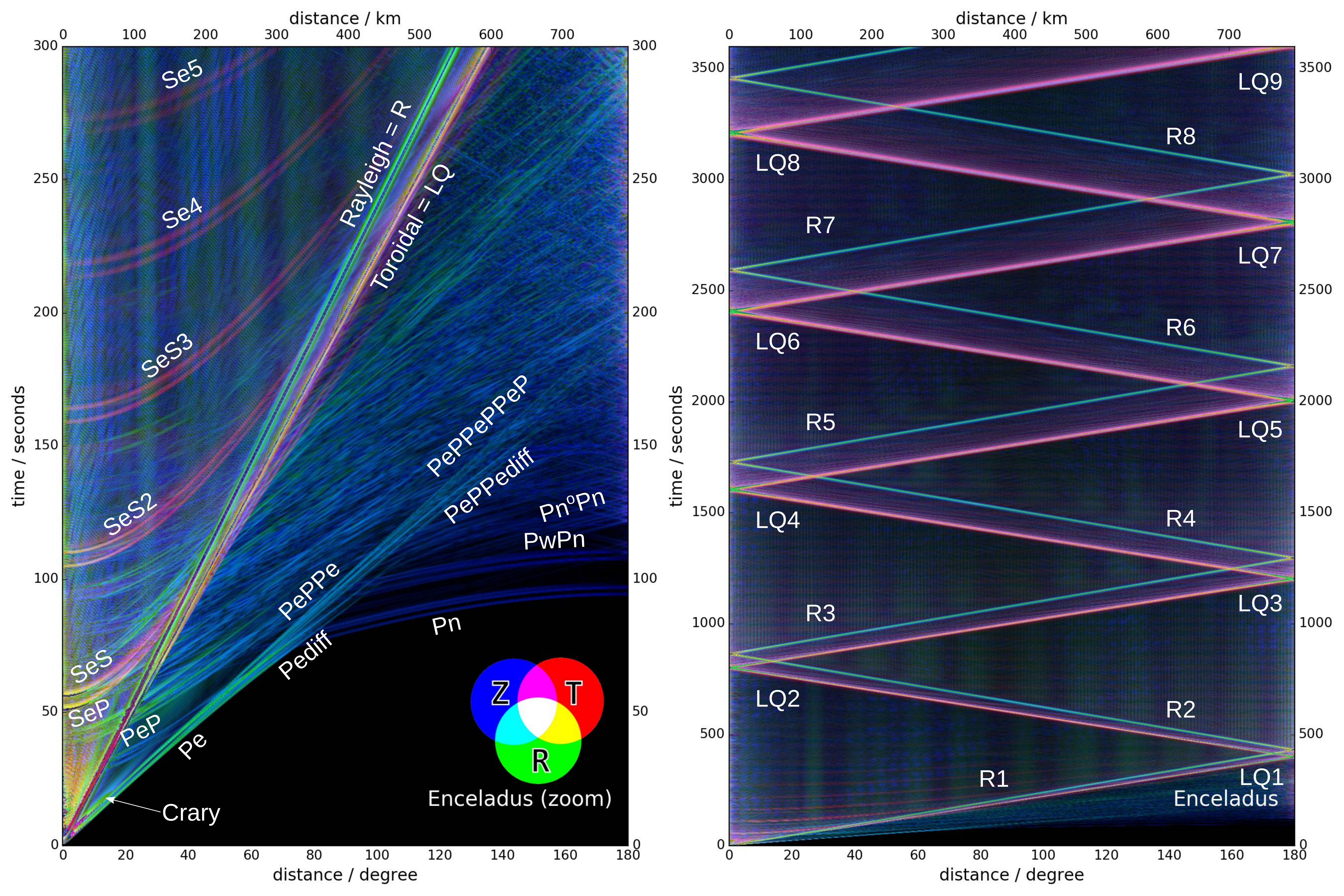}
\caption{Global stack of seismograms for a Enceladus model of 52~km ice thickness. Since this is 20\% of the radius, ice phases (Pe, Pediff are the first to arrive up to 75 degree distance) and surface reflected ice phases are prominent in the coda for all distances. The ice layer is too thick for a measurable longitudinal or Crary phase. Due to the small diameter and the relatively low attenuation, surface waves orbit the moon multiple times and dominate the seismogram (see fig. \ref{fig:comparison_moons}).}
\label{fig:stack_Enceladus}
\end{figure}

Enceladus' global subsurface ocean has been confirmed from its measured shape \citep{McKinnon2015} and libration \citep{Thomas2016}, but the ocean thickness probably varies globally, exceeding 15~km at the south pole due to a thinner ice shell, and less than 10~km everywhere else. For the simulation, we assumed a constant global thickness of 10~km.

Due to its small size, geometrical spreading and attenuation of seismic energy plays a small role on Enceladus. As fig. \ref{fig:comparison_moons} and the global wavefield stack in fig. \ref{fig:stack_Enceladus} shows, even for magnitude 3 events, more than 5 orbits of longitudinal phases are observable, which would allow for easy determination of ice thickness and event locations. Tidal heating from the resonance with Dione suggests a high energy budget for seismicity, consistent with the active tectonic deformation of the South Pole region \citep{Porco2006}. 

The plumes at Enceladus' south pole \citep{Porco2006,Hansen2006} are specifically caused by geysers or cryovolcanism. Since geysers are sources of seismic signals on Earth \citep[e.g.][]{Kedar1998}, the south pole region of Enceladus can be expected to have a very high seismic background noise. At the same time, the subterranean volcanic system means that the seismic velocity structure will be very heterogeneous, which may mean that the concepts of global seismology derived from spherically symmetric models presented here are not applicable.

 In addition to these sources, Enceladus should also regularly generate seismic signals in the ice as a result of tidal flexing \citep{Vance2017b}.  The combination of these sources merits further investigation building on the study of \citet{Panning2017}.
\section{Discussion}
This study gives an overview of general characteristics of icy moon seismology on a global scale, but is in no way exhaustive. An obvious limitation is that only spherical models were studied so far. The above-mentioned methods to determine the ice thickness are likely sensitive to ice undulations \citep[e.g.,][]{Nimmo2007, Lefevre2014}. The characteristic frequency of the Crary wave is sensitive to the average thickness. The Love wave cutoff is sensitive to the minimum thickness, while body wave coda measures the thickness at the location of the source and the receiver. Enceladus for example has an ellipticity of 1/50, and probably a variation in ice and ocean thickness of more than 50\% between the south pole and the equator \citep{Thomas2016}, which will strongly affect propagation, especially of multiply orbiting surface waves. From simulations using cylindrically symmetric models, we found that the longitudinal wave is a very robust feature and multiple orbits are detectable, even with strong heterogeneities in the ice. Undulations of the ice bottom may, however, have a very strong effect on the amplitude and characteristic frequency of the Crary wave and therefore on its effectiveness in constraining ice thickness.

On Europa and Enceladus, the existence of plumes suggests that liquid channels exist in the ice that will strongly scatter seismic waves. Therefore, full-3D simulations will be necessary as a next step. 

Seismic wave propagation on icy ocean worlds is a numerically challenging problem, since it requires high frequencies of up to one Hz to resolve seismic features like the Crary wave for thin ice layers. Since the numerical cost in full 3D methods typically scales with frequency $f$ as $O(f^4)$, this prohibits their application on planet scales. This is problematic, given that the ice waves may be very sensitive to three-dimensional structure, like varying ice thickness. This is rather different from the situation on Earth, were surface waves are strongest at longer periods (typically above 15 seconds), which are possible to simulate on large HPC systems. 
Seismic wavefield solvers suitable for smooth three-dimensional structure at planetary scales at frequencies of up to one Hz are just becoming available \citep{Leng2016} and will offer more insight into the question which phases are usable. Local structure around a lander may strongly affect the measured seismic waveforms, which may be simulated by including local 3D wavefield simulations into global 1D simulations.

This article is intended to set the stage of global seismic wave propagation in icy moons on which more detailed studies of effects of three-dimensional structure can build.

Since the presented analysis of the seismic wavefields is in all likelihood incomplete, we make the Instaseis databases available at \textit{http://instaseis.ethz.ch/icy\_ocean\_worlds/}. The scripts to reproduce the figures are available on the seismo-live website \textit{https://www.seismo-live.org}.

Future research should also explore the parameter space of possible mantle compositions and thermal states. As was shown here, seismology in principle provides tools to constrain both, but detailed studies are necessary for each planetary body, in close collaboration with planetary geophysics (see table \ref{tab:objectives} for a brief summary).

\begin{table}
\begin{tabular}{p{3cm}p{4cm}p{1.5cm}p{3cm}p{2cm}}
Scientific objective & Seismic observable & signal strength & distance range & reference\\ \hline
Tectonic activity & Location of seismic events & strong & global & appendix \ref{app:baz} \\ \hline
Ice thickness & Resonant frequency of Crary phase & strong & global & fig. \ref{fig:Crary_spectra} \\
              & Transition frequency between Rayleigh and flexural surface wave & intermediate & global & fig. \ref{fig:spectrogram}\\
              & Reverberations in body wave coda & strong & teleseismic & fig. \ref{fig:Pcoda}\\
              & Autocorrelation of ambient noise & strong & ambient noise & \citep{Panning2017}\\ \hline
Ocean depth & Reverberations in body wave coda & strong & teleseismic & fig. \ref{fig:Pcoda}\\
            & Autocorrelation of noise & intermediate & ambient noise & \citep{Panning2017}\\ 
Ocean chemistry (from sound velocity in water) & Reverberations in body wave coda & strong & teleseismic & fig. \ref{fig:Pcoda}\\            
\hline
High pressure ice phases & Coda of Sn-waves & strong & teleseismic \\ fig. \ref{fig:Titan_iceVI}\\
                         & Scholte waves & strong & teleseismic  &	\\
                         & P-to S ratio & intermediate & teleseismic, depends on focal mechanism & fig. \ref{fig:Titan_iceVI} \\\hline
Core diameter & Autocorrelation of noise & weak & ambient noise & \citep{Panning2017}\\                          
 & Autocorrelation of seismogram & intermediate & $>100$ deg  & fig. \ref{fig:Pcoda}\\
\end{tabular}
\caption{Potential scientific objectives of a future icy ocean world seismometer and seismic observables that address it.}
\label{tab:objectives}
\end{table}

\section{Conclusion}
In this paper, we discussed the general characteristics of global seismology on icy ocean worlds. The existence of the liquid-filled gap between the mantle and the icy crust creates a different wavefield than the one that is known on earth. In general, this means that a large part of seismic energy remains close to the surface, where it can be measured, which makes even relatively small events observable globally. Also, several seismic measurables are specifically sensitive to parameters related to habitability. This is a promising perspective for the short term installations of seismometers on icy moons.
\section{Acknowledgements}
The authors acknowledge computational support in the project pr63qo "3D wave propagation and rupture: forward and inverse problem" at \textit{Leibniz-Rechenzentrum} Garching. SCS was supported by grant SI1538/4-1 of Deutsche Forschungsgemeinschaft \textit{DFG}. MvD was supported by grants from the Swiss National Science Foundation (SNF-ANR project 157133 "Seismology on Mars") and the Swiss National
Supercomputing Center (CSCS) under project ID sm682.

This work was partially supported by strategic research and technology funds from the Jet Propulsion Laboratory, Caltech, and by the Icy Worlds node of NASA's Astrobiology Institute (13-13NAI7\_2-0024). A part of the research was carried out at the Jet Propulsion Laboratory, California Institute of Technology, under a contract with the National Aeronautics and Space Administration.

The Instaseis wavefield databases available at \textit{http://instaseis.ethz.ch}. The scripts to reproduce the figures are available on the seismo-live website \textit{https://www.seismo-live.org}.

All rights reserved prior to publication.
\pagebreak
\appendix
\section{Determine the backazimuth of an event from a Europan seismogram}
\label{app:baz}
\begin{figure}
\includegraphics[width=0.95\textwidth]{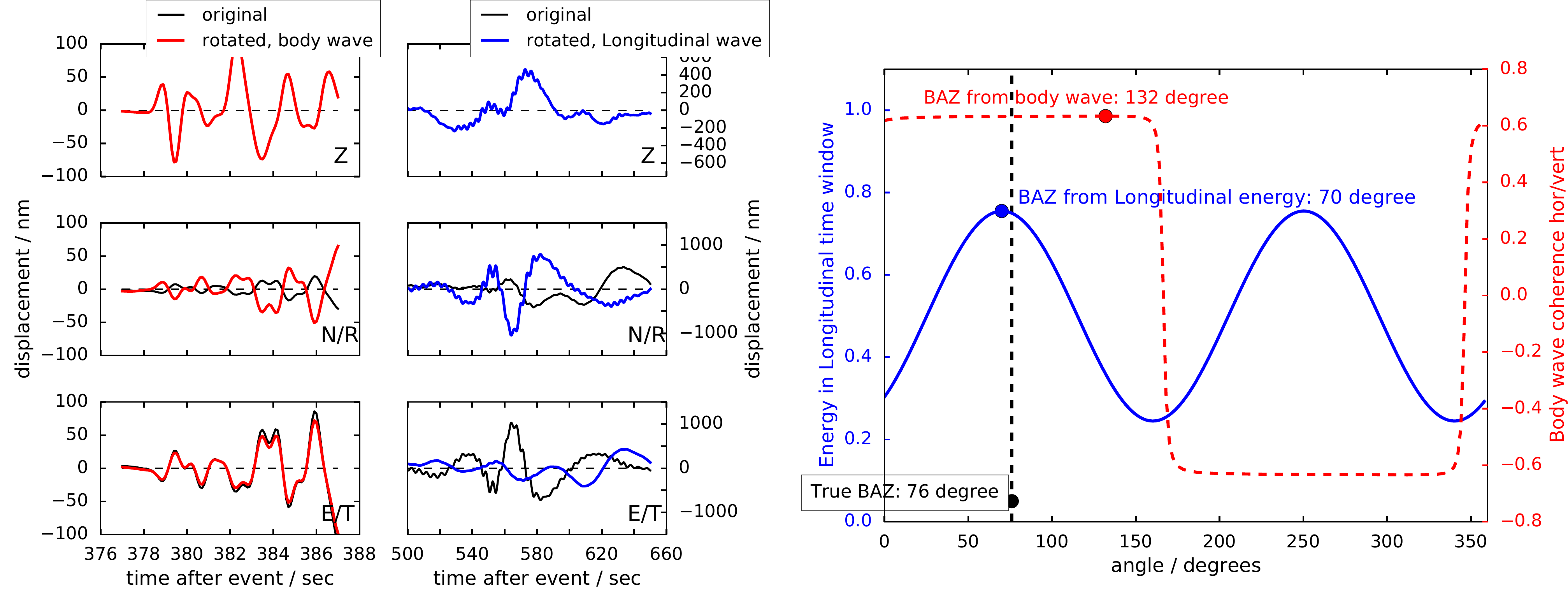}
\caption{Determining the backazimuth with two methods: 1. The energy of the Crary wave should be maximal on the R-component and minimal of T. 2. The first arriving body waves are mainly P-SV waves, therefore coherence between Z and R should be maximized by rotating the seismogram into the correct backazimuth. The seismogram belongs to a M 3.1 event in teleseismic distance with a true backazimuth of 76 degree. The right figure shows the original waveforms in the selected time windows compared to the ones in the optimally rotated system.}
\label{fig:BAZ}
\end{figure}

As described in section \ref{sec:single}, the radial and transverse component of ocean world seismograms differ strongly, which can be used to estimate the backazimuth of an earthquake (i.e. the direction of the earthquake as seen from the receiver). We tested a simple automated implementation of two methods.
\begin{enumerate}
\item Maximize the energy of the Longitudinal wave. The Longitudinal wave is purely longitudinally polarized. Therefore, the horizontal components of the seismogram can be rotated such that energy in the time window around the longitudinal wave is maximal one one horizontal component (R) and minimal on the one perpendicular to it (T). Note that in the presence of noise, the energy in the T-component will not be zero. Also, this method has a 180 degree ambiguity, since the polarity of the longitudinal wave can be positive or negative.
\item Maximize coherence between vertical and radial component for body waves. The first arriving body waves are compressional waves with a high incidence angle, which have motion purely along the direction of propagation (which is a linear combination of the Z-direction and the R-direction pointing away from the event). The direction away from the event is the one, which has maximal correlation to the vertical component. This method has a much lower resolution than the first one, but has no 180 degree ambiguity and so can used to remove that ambiguity from the higher reolution method 1. 
\end{enumerate}
Figure \ref{fig:BAZ} shows the application of both methods to a M3 Europaquake in 83 degree great circle arc distance with a true backazimuth of 76 degree for a Europa model with 5~km ice thickness. It shows that both methods together correctly identify the backazimuth within a few degree.

\section{Properties of the numerical simulations} 
The numerical simulations presented in this manuscript were done using the open source spectral-element wavefield solver AxiSEM. Some modifications were done on the mesher, to make it more stable around very strong velocity contrasts close to the surface. These modifications were added to the main development branch of the software and are included in the recent version 1.4. The simulations were run on the Linux HPC systems \textit{SuperMUC} of the Leibniz-Rechenzentrum Garching and \textit{Piz Daint} of the Swiss National Supercomputing Centre CSCS. 

Due to the low velocity in the ocean, more (small) elements are needed and simulation is about 50\% more expensive compared to a terrestrial planet of the same radius. A 1~Hz simulation for Ganymede or Titan is therefore slightly more expensive than one of the Mars runs done for the InSight blindtest \citep{Ceylan2017, Clinton2017}. 
\label{app:sim_properties}
\begin{table}
\begin{tabular}{|p{1.9cm}|c|r|r|r|r|r|} \hline
Body & Radius & Period & nelement & time step & nsteps & cost\\
 & km & s & & ms & $10^9$ & CPUh \\ \hline
Enceladus & & 0.25 & 1406304 & 2.73 & 1854.5 & 3999\\
 &  & 0.5 & 355224 & 5.6 & 228.4 & 492\\
 &  & 1 & 90630 & 11.4 & 28.6  & 62\\ \hline
Europa  & 1565 & 1 & 2139072 & 8.88 & 867.4  & 1871\\
(5km ice) &  & 2 & 546000 & 15.2 & 129.1 & 278\\
Europa  &  & 1 & 2047690 & 8.95 & 824.0 & 1777\\
(20km ice)	      &  & 2 & 521136 & 15.4 & 121.6 & 262\\\hline
Titan  & 2574 & 1 & 7663680 & 12.6 & 2189.5 & 4722\\
(33km ice)	     &  & 2 & 1929600 & 25.2 & 276.1 & 595\\
Titan  &  & 1 & 7416318 & 9.94 & 2687.3 & 5795\\
(124km ice)             &  & 2 & 1873808 & 19.9 & 339.6 & 732\\\hline
Ganymede & 2631 & 1 & 8281306 & 11.5 & 2584.6  & 5574\\
             &  & 2 & 2090130 & 23 & 326.8  & 705\\\hline
Mars & 3396 & 1 & 5098080 & 7.7 & 2371.5 & 5114 \\ \hline
Earth (PREM) & 6371 & 1 & 9520000 & 7.7 & 4436.8 & 9568 \\ \hline 
\end{tabular}
\caption{Properties of the AxiSEM simulations for this paper. Since AxiSEM is a 2D method, the numerical cost scales roughly with the inverse cube of the minimum period. NCPU is the optimal number of CPUs for mesh decomposition; nelement the number of elements of the SEM mesh. nsteps is the number of elements times the number of time steps for a 1 hour simulation. This is directly proportional to the computational cost, which is shown here on Piz Daint, a Cray XC40 system at the Swiss National Supercomputing Centre CSCS. For comparison, the cost of a 1 hour simulation of Mars (model DWAK in \citet{Clinton2017}) and Earth (PREM) are shown. These costs are based on using gfortran 6.1 and may be lower for more optimized compilers} 
\label{tab:sim_properties}
\end{table}

\section{Retrieving seismic waveforms for ocean worlds}
While the forward simulations of the seismic wavefield require thousands of CPU-hours; the \textit{Instaseis}-database-approach allows to reuse the stored wavefield to calculate seismograms for arbitrary source-receiver locations on the fly \citep{vanDriel2015}. The disk space requirement of the databases depends on the maximum depth of an event, but is generally in the range of several ten Gigabyte. Databases for the models shown in table \ref{tab:sim_properties} are stored on a server at ETH Z\"urich. Similarly to the terrestrial databases stored at the IRIS service \textit{Syngine} \citep{Krischer2017}, they can be openly accessed via the Python package \textit{Instaseis} (http://www.instaseis.net). The following example shows the retrieval of the waveform for a 5.2, pure $M_{rr}$ event in 40 degree distance on Ganymede, including storage into a MiniSEED file.
\begin{verbatim}
import instaseis
db_path = 'http://instaseis.ethz.ch/icy_ocean_worlds/Gan126km-00pMS-hQ_hyd30km_2s'
db = instaseis.open_db(db_path)
src = instaseis.Source(latitude=0.0, longitude=0.0, m_rr=1e17)
rec = instaseis.Receiver(latitude=40, longitude=00.0)
st = db.get_seismograms(src, rec)
st.write('Ganymede_event.mseed', format='MSEED')
\end{verbatim}
%

\end{document}